\documentclass[a4paper,fleqn]{cas-dc}

\usepackage{graphicx}
\usepackage[font=scriptsize, justification=centering]{caption}
\usepackage{subcaption}
\usepackage[linesnumbered, ruled,vlined]{algorithm2e}
\usepackage[switch]{lineno}
\usepackage{multirow}
\usepackage{enumitem}
\usepackage{listings}
\usepackage{xspace}
\usepackage{tcolorbox}
\usepackage{mdframed}
\usepackage[flushleft]{threeparttable}
\usepackage[numbers]{natbib}
\usepackage{xcolor}

\def\tsc#1{\csdef{#1}{\textsc{\lowercase{#1}}\xspace}}
\tsc{WGM}
\tsc{QE}
\tsc{EP}
\tsc{PMS}
\tsc{BEC}
\tsc{DE}

\newdefinition{definition}{Definition}

\definecolor{light-gray}{gray}{0.92} 
\newenvironment{gtheorem}%
{\begin{mdframed}[backgroundcolor=light-gray,
skipabove=5pt,
skipbelow=0pt,
nobreak=false
]\begin{mdtheorem}{name}{label}}%
{\end{mdtheorem}\end{mdframed}}
\definecolor{ao}{rgb}{0.0, 0.5, 0.0}
\newcommand{\tool}{\textsc{AutoProbe}\xspace}
\newcommand{\etal}{\textit{et al.}\space}
\newcommand{\magiccoder}{\textsc{Magicoder}\xspace}
\newcommand{\deepseek}{\textsc{DeepSeek Coder}\xspace}
\newcommand{\codellama}{\textsc{Code Llama}\xspace}
\newcommand{\codegemma}{\textsc{Code Gemma}\xspace}

\newcommand{\gpt}{o4-mini\xspace}
\newcommand{\inhouse}{\textsc{In-house LLM-AJ}\xspace}
\newcommand{\external}{\textsc{External LLM-AJ}\xspace}

\newcommand{\openia}{\textsc{Openia}\xspace}
\newcommand{\oracle}{\textsc{Oracle}\xspace}
\newcommand{\ncor}{\textit{\#Cor.}\xspace}
\newcommand{\ninc}{\textit{\#Inc.}\xspace}

\begin{document}
\let\WriteBookmarks\relax
\def\floatpagepagefraction{1}
\def\textpagefraction{.001}

\shorttitle{\tool}

\shortauthors{Vu \textit{et~al.}}

\title [mode = title]{Model-Agnostic Correctness Assessment for LLM-Generated Code via Dynamic Internal Representation Selection}    

\author{Thanh Trong Vu}
\ead{thanhvu@vnu.edu.vn}
\affiliation{organization={Faculty of Information Technology, VNU University of Engineering and Technology},
    city={Hanoi},
    country={Vietnam}}

\author{Tuan-Dung Bui}
[orcid=0009-0007-7318-6896]
\ead{21020006@vnu.edu.vn}

\author{Thu-Trang Nguyen}
[orcid=0000-0002-3596-2352]
\ead{trang.nguyen@vnu.edu.vn}

\author{Son Nguyen}
[orcid=0000-0002-8970-9870]
\ead{sonnguyen@vnu.edu.vn}
\cormark[1]

\author{Hieu Dinh Vo}
[orcid=0000-0002-9407-1971]
\ead{hieuvd@vnu.edu.vn}

\cortext[cor1]{Corresponding author}


\begin{abstract}
Large Language Models (LLMs) have demonstrated impressive capabilities in code generation and are increasingly integrated into the software development process. 
However, ensuring the correctness of LLM-generated code remains a critical concern. Prior work has shown that the internal representations of LLMs encode meaningful signals for assessing code correctness. Nevertheless, the existing methods rely on representations from pre-selected/fixed layers and token positions, which could limit its generalizability across diverse model architectures and tasks.
In this work, we introduce \tool, a novel model-agnostic approach that dynamically selects the most informative internal representations for code correctness assessment. \tool employs an attention-based mechanism to learn importance scores for hidden states, enabling it to focus on the most relevant features. These weighted representations are then aggregated and passed to a probing classifier to predict code correctness across multiple dimensions, including compilability, functionality, and security.
To evaluate the performance of \tool, we conduct extensive experiments across multiple benchmarks and code LLMs. 
Our experimental results show that \tool consistently outperforms the baselines. For security assessment, \tool surpasses the state-of-the-art white-box approach by 18\%. For compilability and functionality assessment, \tool demonstrates its highest robustness to code complexity, with the performance higher than the other approaches by up to 19\% and 111\%, respectively.   
These findings highlight that dynamically selecting important internal signals enables \tool to serve as a robust and generalizable solution for assessing the correctness of code generated by various LLMs.
\end{abstract}


\maketitle

\section{Introduction}


In recent years, Large Language Models (LLMs) have exhibited their remarkable capabilities in code generation~\cite{HumanEval, rambo, zhang2023repocoder, zheng2025towards} and are increasingly being integrated into the modern software development process~\cite{dai2025comprehensive, github_copilot, codegeex2}. Especially, Code LLMs, such as \deepseek~\cite{deepseek-coder} or \codellama~\cite{codellama}, are fine-tuned on extensive code corpora, enabling them to effectively leverage both natural language understanding and program synthesis abilities. These models have demonstrated strong performance on various code generation benchmarks~\cite{HumanEval, MBPP} and have shown promise in automating a wide range of software engineering (SE) tasks~\cite{zheng2025towards}.

However, recent empirical studies reveal that code generation performance of LLMs often degrades in practical development scenarios~\cite{rambo, zhang2023repocoder, liu2023your, li-etal-2024-deveval}. Although LLMs are capable of generating plausible code, their outputs could be functionally incorrect or insecure~\cite{liu2023your, perry2023users, liu2024no}. 
Perry~\etal~\cite{perry2023users} report that while AI assistants can significantly improve software development productivity, they can also result in more insecure code. Additionally, Tihanyi~\etal~\cite{tihanyi2025secure} employed SMT-based Context-Bounded Model Checker to validate LLM-generated code and found that over 62\% of the generated programs contain vulnerabilities.

As LLM-generated code is increasingly integrated into real-world applications, ensuring its correctness and reliability has become critical~\cite{msr2023-prem-llm-code-bugs,llm-code-bugs1,code-quality-chagpt,evaluating-chatgpt}. 
Low-quality code can lead to significant functional failures, security flaws, and increased technical debt. These risks demand effective and rigorous evaluation techniques to assess the quality and reliability of code produced by LLMs~\cite{survey_3, survey-icse24, expectation}.

To address the challenge of assessing the correctness of LLM-generated code, recent studies have explored both \textit{black-box} and \textit{white-box} strategies. Black-box approaches~\cite{embedding_emse22,opt-pretrained-model,pretrained-survey} typically rely solely on the final generated outputs. For example, pre-trained models such as CodeBERT~\cite{codebert} and CodeT5~\cite{codet5} are employed to encode the semantics of the LLM-generated code, and a downstream classifier then predicts whether the code is correct. Meanwhile, white-box approaches~\cite{openia, internal-state-2, inner-working, probing, internal-state, inside} leverage the internal reasoning process of the LLM during generations, e.g., hidden states or attention activations, to detect correctness. These methods have demonstrated that internal representations of LLMs encode meaningful signals related to code correctness, enabling the white-box approaches to outperform the black-box ones~\cite{openia, internal-state-2, internal-state} across multiple benchmarks.

Despite their potential performance, these existing white-box approaches~\cite{openia, internal-state-2, internal-state} adopt a \textit{rigid representation selection strategy}, which restricts their generalizability and robustness. 
These approaches assume that correctness-related signals consistently reside in the same layers and token positions. However, in practice, the distribution of these signals is highly dependent on model architectures and tasks. 
As a result, relying on a \textit{fixed} selection of layer and token positions can lead to the omission of critical information and result in suboptimal performance. 
Moreover, since there is no one-size-fits-all strategy across LLMs, a hard-coded selection strategy is less adaptable to the diversity of the models, reducing its effectiveness in real-world scenarios.

In this work, we introduce \tool, a novel \textit{model-agnostic white-box} framework for assessing the correctness of LLM-generated code. Our key insight is that not all hidden states contribute equally to correctness prediction, and the positions of important signals vary across LLM architectures. To effectively capture the most informative signals, \tool employs an attention-based mechanism to learn importance scores over all internal states. This process allows a probing classifier to focus on the most relevant features while down-weighting the irrelevant ones. The weighted representations are then aggregated and passed to a probing classifier to determine whether the generated code is correct. By \textit{dynamically selecting meaningful signals}, \tool improves its robustness and generalizability across diverse LLMs.

To evaluate the performance of \tool, we conduct extensive experiments across multiple benchmarks and a set of six code LLMs. The experimental results demonstrate that \tool consistently outperforms  baseline approaches across all evaluation dimensions. In security assessment, \tool surpasses the state-of-the-art (SOTA) white-box method, \openia, by up to 18\% in accuracy. For compilability and functionality assessment, it exhibits exceptional robustness to increased code complexity, achieving performance gains of up to 19\% and 111\%, respectively. These findings highlight \tool’s capability to deliver accurate and reliable correctness assessment across varied tasks, models, and complexity levels.

In brief, this paper makes the following contributions:
\begin{itemize}
  \item We formalize the task of code correctness assessment across multiple dimensions, including  \textit{compilability}, \textit{functionality}, and \textit{security}, providing a rigorous foundation for evaluating LLM-generated code.
   \item We propose \tool, a novel model-agnostic framework that dynamically selects informative internal representations of LLMs during generations for assessing code correctness.
    \item We conduct extensive experiment evaluations demonstrating the robustness and generalizability of \tool across diverse code LLMs and benchmarks.
\end{itemize}

\section{Background and Motivation}
\subsection{Background}

Large Language Models (LLMs), such as GPT-3~\cite{floridi2020gpt}, PaLM~\cite{anil2023palm}, and LlaMA~\cite{touvron2023llama}, are typically transformer-based models with hundreds of millions to billions of parameters~\cite{minaee2024large}. These models are pre-trained on massive text corpora using an autoregressive objective, i.e., predicting the next token given the preceding context, to model the probability distribution over token sequences.
This training paradigm enables LLMs to understand and generate coherent text, allowing them to perform a wide range of natural language tasks, like translation~\cite{ji2024zero}, summarization~\cite{zhang2025systematic}, and question answering~\cite{li2024flexkbqa}.
To extend their capabilities beyond natural language, LLMs have been trained or fine-tuned on large-scale code repositories, giving rise to a specialized class known as \textbf{code LLMs}, such as \deepseek~\cite{deepseek-coder}, \magiccoder~\cite{magicoder}, \codellama~\cite{codellama}.
These models are capable of automating various SE tasks~\cite{zheng2025towards, wang2025can, bui2024rambo, chen2024chatunitest}, including code generation and code completion.

Similar to general-purpose LLMs, code LLMs generate code output through an autoregressive decoding procedure. Given an input sequence $x = \{x_1, \dots, x_n\}$, which may consist of natural language instructions, code comments, and/or partially written code, the code LLM $\mathscr{M}$ produces a code output sequence $c = \{c_1, \dots, c_m\}$ one token at a time. At each generation step $s$, the model computes the probability distribution over possible next token $c_s$, conditioned on the input $x$ and previously generated tokens $c_{<s>} = \{c_1, \dots, c_{s-1}\}$, as follows:
\begin{equation}
\label{eq:next_token_probability}
 P(c_s \mid x,c_{<s}) = \text{softmax}(W_o \cdot h_{l,s} + b_o)   
\end{equation}
In Eq.~\ref{eq:next_token_probability}, $h_{l,s} \in \mathbb{R}^d$ denotes the  \textit{internal representation} (or hidden state) of the model at layer $l$ for the token position $s$, and $W_o$, $b_o$ are learned output projection weights.

\begin{definition}[\textbf{Internal representation}]
An internal representation $h_{l,s} \in \mathbb{R}^d$ is a $d$-dimensional vector that encodes the model's latent representation for token $c_{s}$ at layer $l$.   
It is recursively computed from the hidden state at the previous layer $h_{l-1, s}$ and contextual information  $context_s$ via a layer-specific transformation function $f_l(\cdot)$:
\begin{equation}
\label{eq:hidden_state_computation}
    h_{l,s} = f_l(h_{l-1,s}, \text{$context$}_s)
\end{equation}

\end{definition}

The transformation function $f_l(\cdot)$ in Eq.~\ref{eq:hidden_state_computation} typically consists of a combination of multi-head self-attention, feedforward networks, and normalization operations.
The internal representations form the core computational states of the model, capturing essential information at different levels of abstraction, including syntactic structure, semantic meaning, and task-specific context, that are required for token prediction at each generation step.

Despite their impressive capabilities, LLMs, including code LLMs, are prone to \textit{hallucinations}. In the context of code generation, hallucinations refer to code that is non-compilable, functionally incorrect, or insecure, potentially causing the program to crash, fail to meet requirements, or become vulnerable to attacks~\cite{openia, zhang2025llm}. 
Several studies~\cite{openia, llmcheck, inside, zhang2025icr} have demonstrated that the model's internal representations encode signals indicative of hallucinations. Since these internal states reflect the model's evolving understanding and confidence during generation, they can serve as valuable indicators of uncertainty or inconsistency. By analyzing the internal representations, it is possible to detect hallucinations both in text and code generation with a certain degree of accuracy.

\subsection{Code Correctness Assessment Problem Formulation}
\label{sec:problem_formulation}

Given the prevalence of hallucinations in LLM-generated code and the potential of internal representations to signal such failures, we aim to formulate the task of \textit{code correctness assessment} based on a model's internal states.
Let $\mathscr{M}$ be a code LLM consisting of $L$ layer, which generates sequence of code tokens $c = \{c_1, \dots, c_m\}$ given an input $x = \{x_1, \dots, x_n\}$. At each generation step $s$, $s \in [1, m]$, the model produces internal representation $h_{l,s}$ at every layer $l$, $l \in [1, L]$, capturing contextualized information about token $c_s$. We define $\mathcal{H} = \{ h_{l,s} \mid l \in [1, L],\ s \in [1, m] \}$ as the set of internal representations associated with the entire generated sequence, across all $L$ layers and $m$ token positions.

Formally, we define the \textit{code correctness assessment} task as a binary classification problem:

\begin{definition}[\textbf{Code Correctness Assessment}]
Given the internal representations $\mathcal{H}$ associated with the generated code sequence $c$, the goal of code correctness assessment is to predict whether $c$ is correct or not. Formally,  we aim to learn a function:

$$
f_\theta: \mathcal{H} \mapsto y
$$
where $f_\theta$ maps the internal states $\mathcal{H}$ to a correctness label $y \in \{0, 1\}$, where $y = 1$ indicating correct code and $y = 0$ indicating incorrect or hallucinated code.
\end{definition}

The correctness of the generated code can be evaluated along multiple dimensions, including \textit{compilability}, \textit{functionality}, and \textit{security}. Depending on the use case, different oracles, such as compilers, dynamic test cases, or static security analyzers, can be employed to provide ground-truth correctness labels.
 
\begin{definition}[\textbf{Compilability}]

Compilability refers to whether the generated code $c$ can be parsed and compiled successfully without syntax or structural errors.  
Let $\texttt{compile}(c)$ denote the result of compiling $c$ using a standard compiler (e.g., \texttt{gcc} for C/C++, \texttt{javac} for Java) , where $\texttt{compile}(c) \in \{\textit{success}, \textit{failure}\}$.  
Code $c$ is considered compilable if and only if it compiles without errors:
\[
\chi(c) =
\begin{cases}
1 & \text{if } \texttt{compile}(c) = \textit{success}, \\
0 & \text{otherwise}.
\end{cases}
\]
\end{definition}

\begin{definition}[\textbf{Functionality}]
Functionality ensures that the generated code $c$ satisfies the functional requirements specified by the input prompt $x$. 
Let $\mathcal{T} = \{t_1, \dots, t_k\}$ denote a set of functional test cases designed to validate whether $c$ fulfills the intended behaviors. Let $t(c)$ represent the outcome of applying a test $t \in \mathcal{T}$ to the code $c$, where $t(c) \in \{passed, failed\}$.
Code $c$ is considered functionally correct if and only if it passes all tests in  $\mathcal{T}$: 

\begin{equation}
\eta(c) = 
\begin{cases} 
1 & \text{if } \forall t \in \mathcal{T}, \; t(c) = \textit{passed}, \\ 
0 & \text{otherwise}.
\end{cases}
\end{equation}

\end{definition}

\begin{definition}[\textbf{Security}]

Security assesses whether the generated code $c$ adheres to secure coding practices. 
Let $\mathcal{V} = \{v_1, \dots, v_h\}$ denote a set of security vulnerability checks, and $v(c)$ represents the outcome of evaluating $c$ against vulnerability check $v \in \mathcal{V}$, where $v(c) \in \{\textit{safe}, \textit{unsafe}\}$.
Code $c$ is considered secure if and only if it passes all security checks in $\mathcal{V}$:

\begin{equation}
\sigma(c) = 
\begin{cases} 
1 & \text{if } \forall v \in \mathcal{V}, \; v(c) = \textit{safe}, \\ 
0 & \text{otherwise}.
\end{cases}    
\end{equation}

\end{definition}

This work focuses on three key dimensions of code correctness: compilability, functionality, and security, as they represent fundamental aspects of practical code quality~\cite{al2011software}. These dimensions cover syntactic validity, behavioral correctness, and safety, each critical for ensuring reliable and trustworthy code. While our framework is tailored to these dimensions, it can be readily extended to incorporate additional correctness criteria in a similar manner.
\subsection{Motivation}

\begin{figure}
\centering
\begin{subfigure}{\columnwidth}
\centering
\includegraphics[width=1\columnwidth]{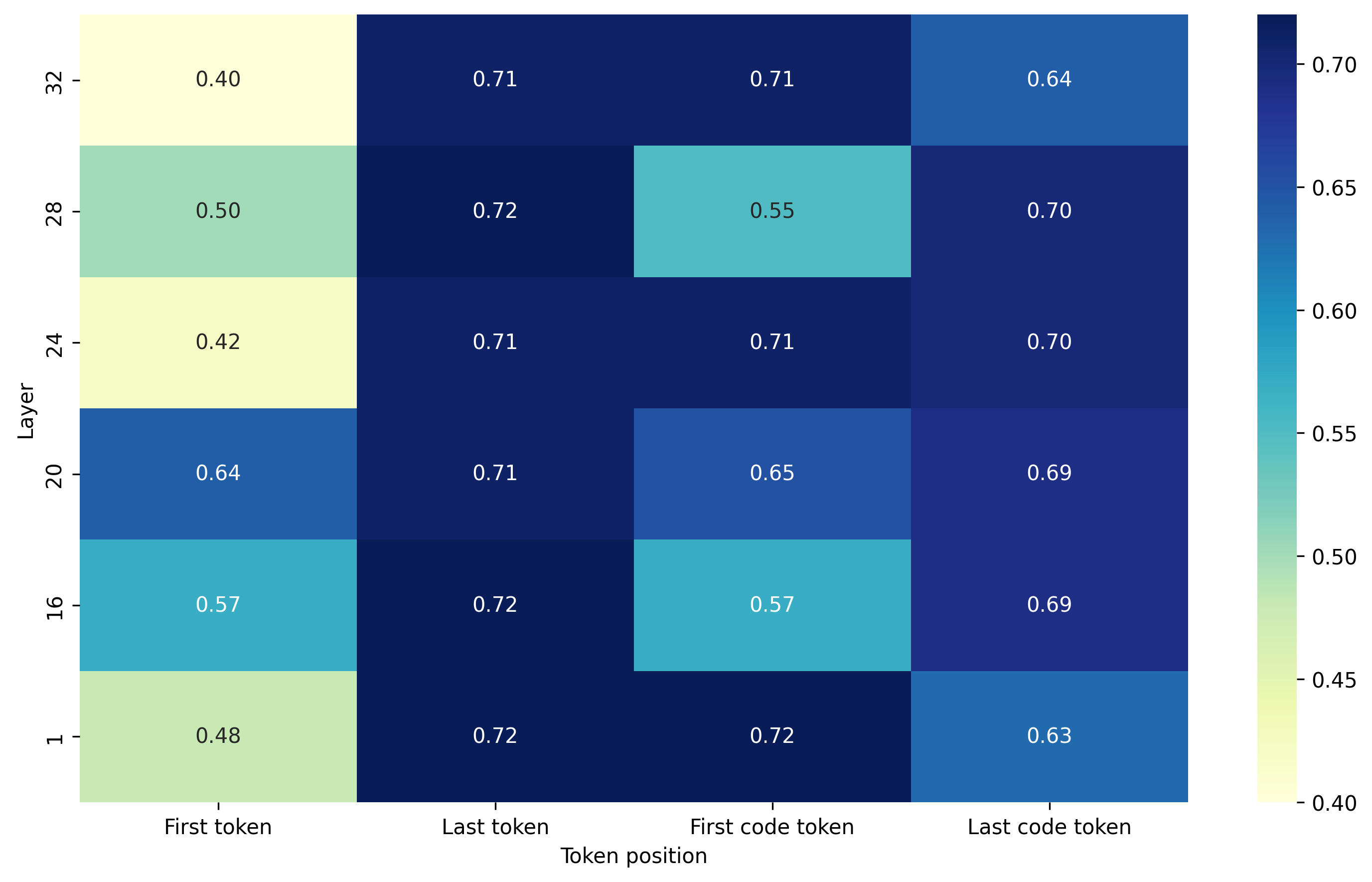}
\caption{\deepseek}
\label{fig:motivation_deepseek}
\end{subfigure}\\
\begin{subfigure}{\columnwidth}
\centering
\includegraphics[width=1\columnwidth]{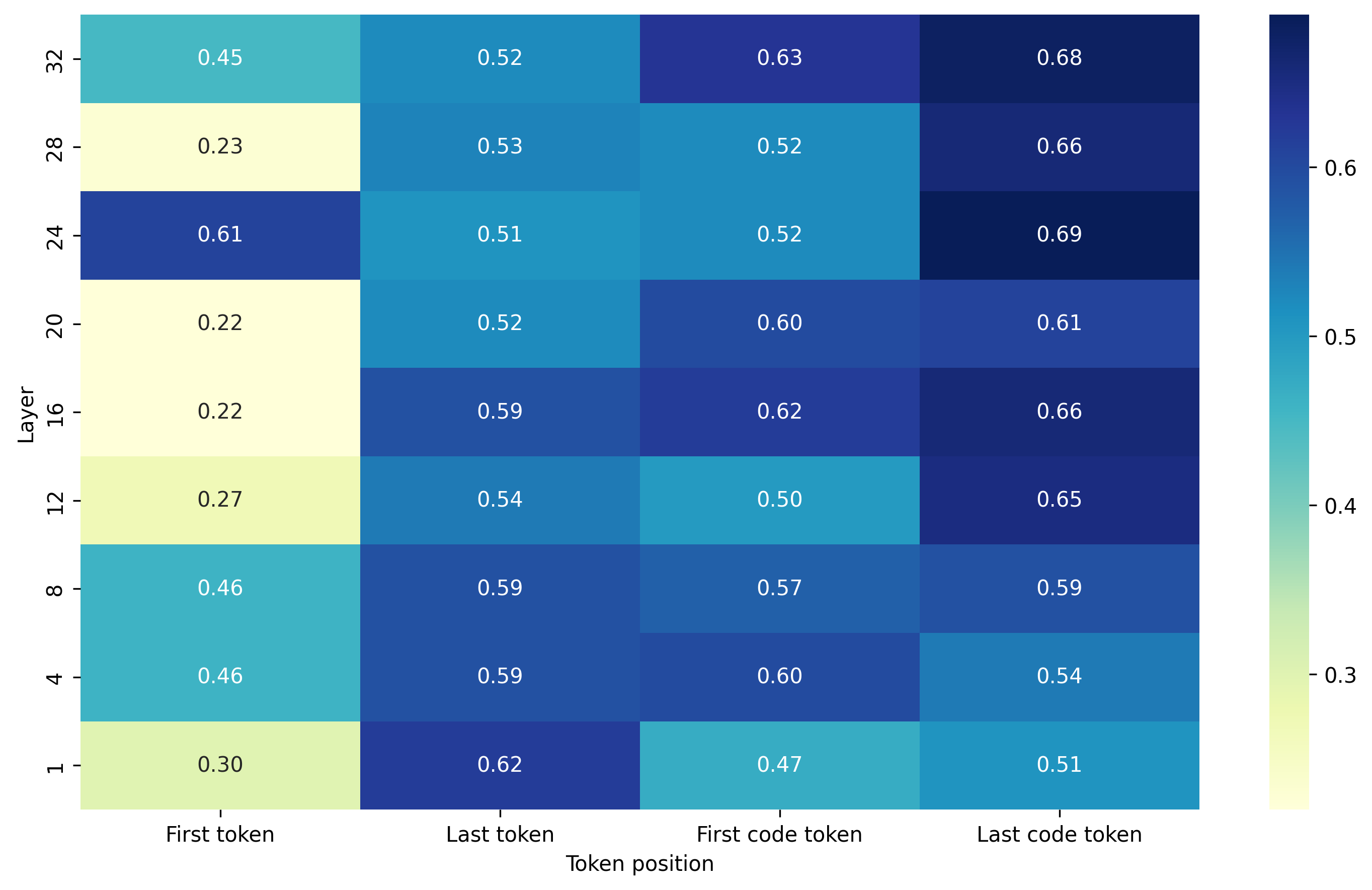}
\caption{\codellama}
\label{fig:motivation_codellama}
\end{subfigure}\\
\begin{subfigure}{\columnwidth}
\centering
\includegraphics[width=1\columnwidth]{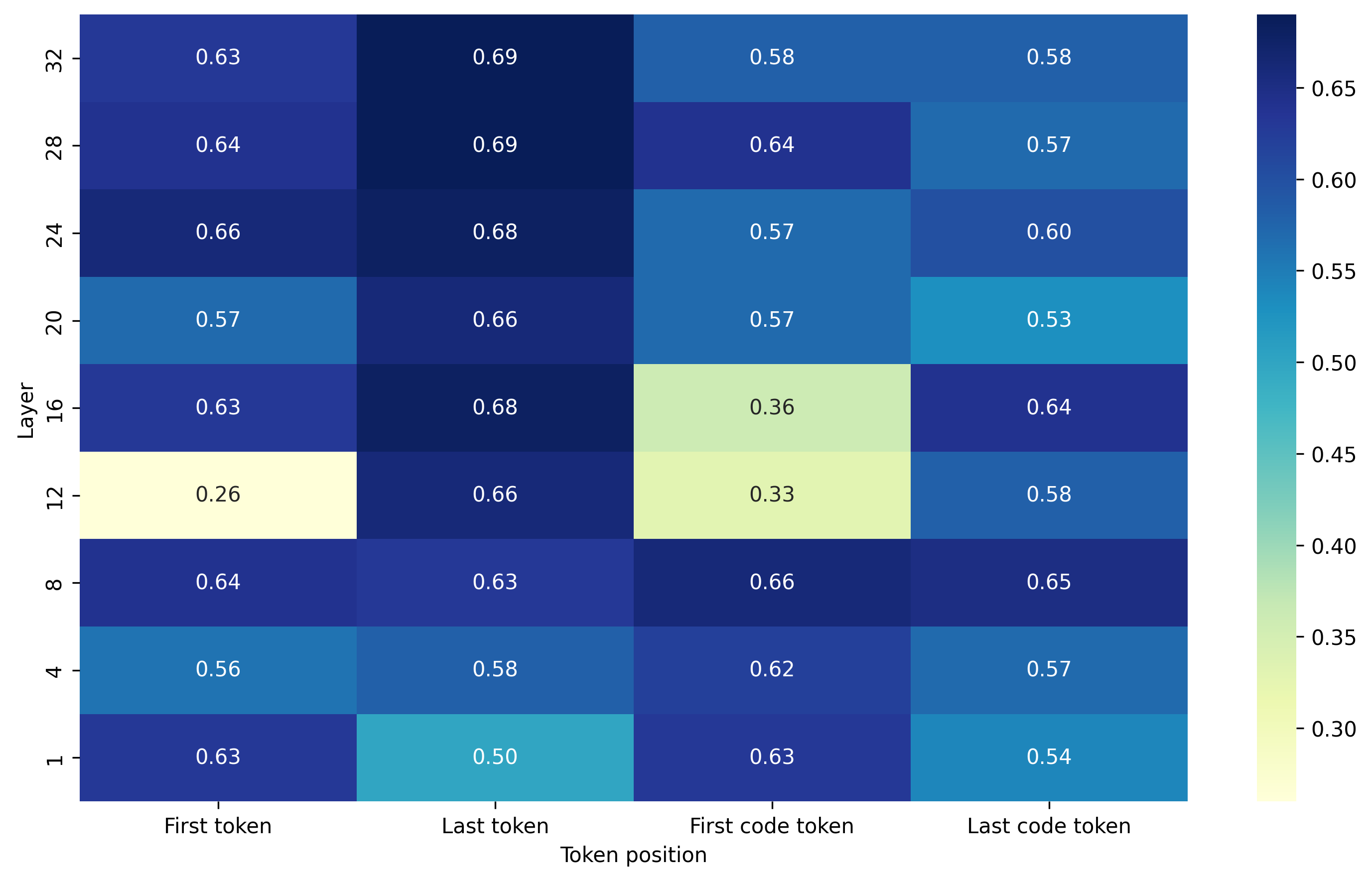}
\caption{\magiccoder}
\label{fig:motivation_magiccoder}
\end{subfigure}

\caption{The performance in F1-score of functionality assessment using internal states from different tokens and layers across various code LLMs}
\label{fig:motivation}
\end{figure}

Inspired by the procedure in \openia~\cite{openia}, we conducted an empirical study to assess the functional correctness of code generated by \deepseek-6.7B, \codellama-7B, and \magiccoder-7B using their internal representations. 
For HumanEval~\cite{HumanEval} and MBPP~\cite{MBPP} benchmarks, we employed each studied model to generate 10 solutions per task.
The functional correctness of each solution is determined by using the provided test cases; these resulting datasets are summarized in Table~\ref{tab:preliminary_data_summarize}. 

\begin{table}\centering
\caption{{The numbers of functional correct (\ncor) and incorrect (\ninc) solutions of benchmarks}}
\label{tab:preliminary_data_summarize}
\scriptsize
\begin{tabular}{l|l|r|r}\toprule
\textbf{Code LLM}&                                               &\textit{HumanEval}  &\textit{MBPP}    \\\midrule

\multirow{2}{*}{\textit{\textbf{\deepseek-6.7B}}}  &\ncor      &1,294      &3,033    \\
                                                  &\ninc      &346        &1,749    \\\midrule

\multirow{2}{*}{\textit{\textbf{\codellama-7B}}}   &\ncor      &644        &1,942    \\
                                                  &\ninc     &996        &2,686    \\\midrule                                               
\multirow{2}{*}{\textit{\textbf{\magiccoder-7B}}}  &\ncor      &1,198      &3,135    \\
                                                  &\ninc     &440        &1,752    \\
\bottomrule
\end{tabular}
\end{table}

For each code LLM, we trained a probing classifier on its internal representations extracted from different layers and token positions during code generation for MBPP tasks. 
The classifier was subsequently evaluated on its ability to predict the correctness of the solutions for HumanEval tasks. In this experiment, representations were extracted from four specific token positions: \texttt{first token} and \texttt{last token} (marking the start and end of the full generated sequence), and \texttt{first code token} and \texttt{last code token} (marking the start and end of the code snippet itself).

Figure~\ref{fig:motivation} shows the performance of probing classifiers using internal representations from various layers and token positions for the studied model. 
The results reveal a key challenge: \textbf{\textit{the code correctness assessment performance varies significantly across models, layers, and token positions}}. 

First, the choice of token position plays a crucial role in the predictive performance of internal representations, with notable variation across models.
For instance, in \deepseek and \magiccoder, the highest performance is achieved using the internal representations of the \texttt{last token}, while in \codellama, the best results are observed from the hidden states of the \texttt{last code token}. Conversely, the representations at \texttt{first token}, while being the least informative for both \deepseek and \codellama, provide comparatively strong signals in \magiccoder.

Moreover, the predictive performance varies significantly across layers, even for the same token position. 
In \magiccoder, the performance improves with layer depth, i.e., the F1-score for the \texttt{last token} at the \texttt{final layer} reaches 0.69, which is 40\% higher than that at the \texttt{first layer}. 
In contrast, \deepseek exhibits remarkably stable performance across layers, maintaining a high F1-score for the \texttt{last token} representations.
\codellama shows highly inconsistent patterns. For the \texttt{last token}, better performance is observed at \texttt{early and middle layers}, while for the \texttt{last code token}, the best F1-scores are achieved using representations from the \texttt{middle and final layers}.

These findings suggest that the optimal internal representations for code correctness assessment are highly model-dependent, with the most effective layers and token positions varying significantly across models.
In other words, \textit{no single layer or token position consistently captures correctness signals across all models}.
This highlights the need for a general solution that can \textbf{\textit{dynamically and automatically select the most informative internal states}}, enabling reliable correctness assessment in a model-agnostic manner.

\section{Model-Agnostic Correctness Assessment for LLM-Generated Code via Dynamic Internal Representation Selection}

\begin{figure*}
    \centering
    \includegraphics[width=\linewidth]{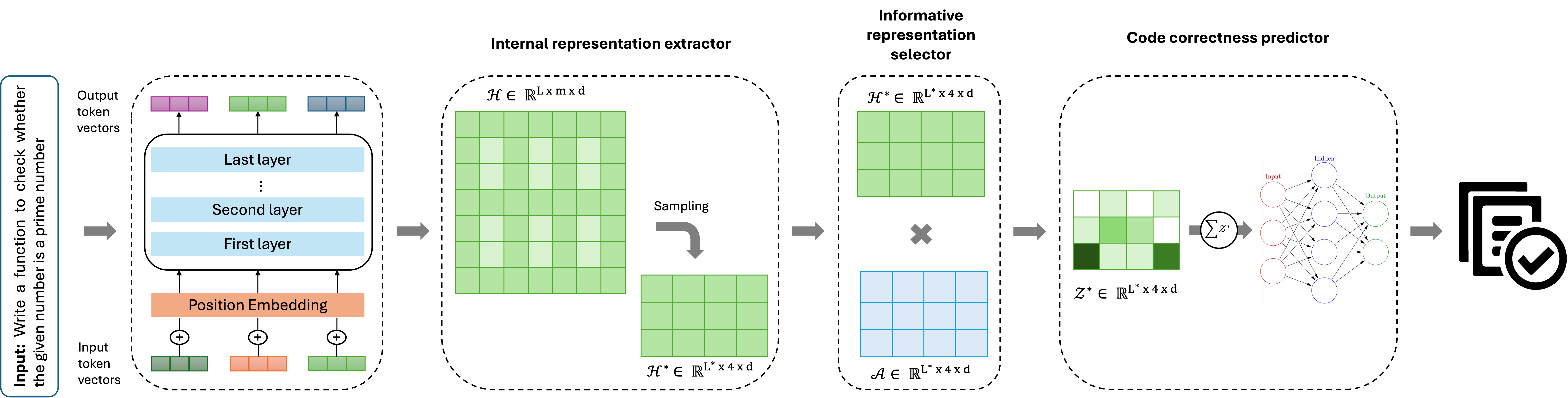}
    \caption{Approach overview}
    \label{fig:approach_overview}
\end{figure*}

\subsection{Approach Overview}
This paper proposes \tool, a novel approach for assessing the correctness of LLM-generated code by dynamically identifying the most informative signals from LLMs' internal states during code generation.
Our key idea is that instead of manually engineering a ``one-size-fits-all'' heuristic, we design a model-agnostic approach that learns to select the most relevant internal representations for code correctness assessment. 
By leveraging a data-driven attention-based mechanism, \tool is able to adapt to the idiosyncrasies of different LLMs, enabling robust and generalizable evaluation across models.
As shown in Figure~\ref{fig:approach_overview}, \tool comprises three key components: \textit{Internal Representation Extractor}, which gathers hidden states from multiple layers and token positions; \textit{Informative Representation Selector}, which prioritizes the most relevant representations based on the attention mechanism; and \textit{Code Correctness Predictor}, which employs a probing classifier to determine whether the generated code is correct based on the dynamically selected features.
\subsection{Internal Representation Extractor}
\label{sec:extractor}

This module collects internal representations produced during code generation as inputs for subsequent components of \tool.
However, the comprehensive set of $L \times m$ internal representations for the entire code sequence $c$ length $m$, $\mathcal{H} = \{ h_{l,s} \mid l \in [1, L],\ s \in [1, m] \}$ is computationally expensive to process and potentially noisy for code correctness assessment. 
Moreover, as illustrated in our empirical study, not all tokens and layers contribute equally to the correctness assessment,  depending on both the model and the generation context.
A naive approach using all representations would therefore be inefficient and could degrade performance.

To improve computational efficiency without sacrificing predictive performance, \textit{Internal Representation Extractor} adopts a sampling-based strategy that extracts a compact yet expressive subset of states $\mathcal{H}^* \subset \mathcal{H}$. This is achieved via a two-pronged strategy that selects representations based on their token and layer positions.

\textbf{Token Position Sampling}. 
For token sampling, rather than uniformly sampling tokens or relying on fixed intervals, we adopt a \textit{boundary-aware token selection heuristic} guided by both the structural and semantic considerations.  Specifically, we select four \textit{structurally and semantically significant} positions: the \texttt{first} and \texttt{last tokens} of the entire generated sequence, and the \texttt{first} and \texttt{last code tokens} that mark the actual code snippet. 
The \texttt{first} and \texttt{last} tokens often reflect the model's overall generation context, as well as its uncertainty or confidence~\cite{yin2024characterizing, snyder2024early}. 
Meanwhile, the \texttt{first} and \texttt{last} code tokens typically carry essential information related to syntax, control flow, and output structure~\cite{openia}.

In practice, token positions can also be selected using other strategies such as syntax-aware analysis, rule-based heuristics, or model-specific indicators. However, we employ this boundary-aware heuristic for several reasons. 
First, prior studies have shown that boundary tokens often carry stronger correctness-related signals than interior positions~\cite{snyder2024early}.
Second, this strategy yields a fixed number of token positions regardless of the sequence length, which varies significantly across tasks. This consistency simplifies downstream processing and model training. 
Third, the heuristic is lightweight, model-agnostic, and adaptable to different code formats.

\textbf{Layer Sampling}. 
To avoid redundantly extracting hidden states from all $L$ layers, we apply a uniform sampling strategy over layers.
Specifically, we sample layers at a fixed interval $k$, starting from the first layer. That is, we select layers at positions $l \in \{1, 1 + k, \dots, 1 + \lfloor \frac{L-1}{k}\rfloor . k\}$.
This results in a total of $L^*$ = $\lfloor \frac{L}{k} \rfloor$ sampled layers. This sampling balances the trade-off between representation diversity and computational complexity.

By restricting the extraction to a compact set of $L^*$ layers and four token positions, we obtain a compact yet expressive set of internal states $\mathcal{H}^* \subset \mathcal{H}$. $\mathcal{H}^*$ serves as input to the subsequent selector module, which identifies the most informative states for final prediction.
Although token and layer sampling may introduce the risk of omitting potentially useful internal states, our empirical results in Sec.~\ref{sec:impact_sampling} demonstrate that this strategy achieves strong predictive performance while substantially reducing computational overhead. These findings validate the effectiveness and efficiency of our sampling approach in practice.

\subsection{Informative Representation Selector}

To identify the most informative internal states within the compact set $\mathcal{H}^*$, \tool employs the \textit{attention-based mechanism} that learns to dynamically prioritize representations based on their relevance to code correctness assessment. 
Particularly, in \textit{Informative Representation Selector}, the attention mechanism assigns a learned importance score to each internal state, quantifying its contribution to the final prediction.
By aggregating the internal states weighted by these scores, the correctness predictor can emphasize the most informative signals while effectively down-weighting irrelevant ones, enabling general and robust assessment across diverse code LLMs.

Specifically, the input of the informative representation selector module is $\mathcal{H}^* = \{h_{l,i} \mid l \in \{1, 1 + k, \dots, 1 + \lfloor \frac{L-1}{k}\rfloor . k\}, i \in {S}\}$, where $S$ denotes the set of four selected token positions, $S =$ \{\texttt{first token}, \texttt{last token}, \texttt{first code token}, \texttt{last code token}\}. 
The selector module assigns a weighted score (attention score) $\alpha_{l,i}$ to each internal state $h_{l,i}$, indicating its importance for correctness prediction. The weighted representation is then computed as:
\begin{equation}
    z_{l,i} = \alpha_{l,i} \cdot h_{l,i}
\end{equation}
The collection of all the weighted vectors $z_{l,i}$ forms the output set $\mathcal{Z}^*$, which summarizes the weighted contributions of the internal states in $\mathcal{H}^*$ to the final correctness prediction. 

To compute attention scores  $\alpha_{l,i}$, each internal state in $\mathcal{H}^*$ is linearly projected and normalized via softmax:
\begin{equation}
\label{eq:attention_score}
    \alpha_{l,i} = \text{softmax}(H \cdot W_a + b_a)_{l,i}
\end{equation}
In Eq.~\ref{eq:attention_score}, $H \in \mathbb{R}^{(L^* \times 4) \times d}$ is the flattened matrix formed by stacking all vectors in $\mathcal{H}^*$. The parameters $W_a \in \mathbb{R}^{d \times 1}$ and $b_a \in \mathbb{R}$ are learnable weights that define the attention scoring function.  The softmax ensures that attention scores across all positions sum to $1$, enabling a normalized weighting over the entire set $\mathcal{H}^*$, i.e., $\sum_{l, i}\alpha_{l,i} = 1$. All the attention scores form an attention sets $\mathcal{A} = \{a_{l,i}\}$.

In \tool, the parameters $W_a$ and $b_a$ are jointly trained with the code correctness predictor using supervised learning. During training, the model receives examples of LLM-generated code along with their correctness labels, which are assigned according the evaluation criteria, e.g., compilability or functionality (Sec.~\ref{sec:problem_formulation}). The cross-entropy loss is measured and back-propagated through the attention mechanism, enabling the model to learn which internal states are most indicative of correctness.

\subsection{Code Correctness Predictor}
\label{sec:predictor}

In \textit{Code Correctness Predictor},
\tool transforms the set weighted representations  $\mathcal{Z}^* = \{z_{l,i}\}$ into a single, fixed-length vector of representation. Specifically, it aggregates all the weighted representations in $\mathcal{Z}^*$ into a unified vector $z \in \mathbb{R}^d$ as follows:
\begin{equation}
\label{eq:representation_aggregation}
    z = \text{aggregate}(\mathcal{Z}^*)
\end{equation}
Without loss of generality, the aggregation function $aggregate$ in Eq.~\ref{eq:representation_aggregation} could be implemented using various strategies such as concatenation, summation, mean/max pooling, etc. The impact of different aggregation methods is empirically evaluated in Sec.~\ref{sec:impact_aggregation_classifier}. 

The aggregated vector $z$ is then fed into a probing classifier that predicts the correctness label of the corresponding code $c$. The probing classifier can be implemented as either a simple neural network or a linear model.
It is trained end-to-end with the selector using standard supervised learning. 
The influence of different classifier architectures on prediction performance is also evaluated in Sec.~\ref{sec:impact_aggregation_classifier}.

\section{Evaluation Methodology}

To evaluate the performance of \tool, we seek to answer the following research questions:

\noindent \textbf{RQ1. Performance Analysis:} How effective is \tool in assessing the correctness of code generated by different code LLMs compared to SOTA methods?

\noindent \textbf{RQ2. Intrinsic Analysis:} How do different components of \tool contribute to its overall performance?

\noindent \textbf{RQ3. Sensitivity Analysis:}  How sensitive is \tool's performance to different external factors such as programming languages, code LLMs, and training data sizes?

\noindent \textbf{RQ4. Efficiency Analysis:} How is \tool's time complexity and memory usage?

\subsection{Dataset Construction}

\label{sec:dataset_construction}
To evaluate the performance of \tool and the baselines, we construct an experimental dataset from several popular benchmarks, including \textit{HumanEval}~\cite{HumanEval}, \textit{MBPP}~\cite{MBPP}, \textit{DevEval}~\cite{li-etal-2024-deveval}, \textit{SecurityEval}~\cite{SecurityEval}, \textit{CWEval}~\cite{Cweval}, \textit{CODEGUARD+}~\cite{CODEGUARD+}, and \textit{SALLM}~\cite{Sallm}.
For each task in these benchmarks, we employed each studied code LLMs to generate 10 candidate solutions.
 
In particular, HumanEval, MBPP, and DevEval focus on functional correctness and are therefore used to construct datasets for evaluating compilability and functionality. The remaining benchmarks are specifically designed to assess security correctness, focusing on identifying potential vulnerabilities in generated code.
Following the procedures established in the prior work~\cite{openia}, the benchmark guidelines, and our criteria defined in Sec.\ref{sec:problem_formulation}, we assign correctness labels to each generated solution as follows:
\begin{itemize}
    \item \textit{Compilability}: A solution is labeled \textit{correct} if it is successfully executed by the Python 3 interpreter; otherwise, it is labeled \textit{incorrect}.
    \item \textit{Functionality}: A solution is labeled \textit{correct} if it passes all provided test cases for the corresponding task; otherwise, \textit{incorrect}.
    \item \textit{Security}: A solution is labeled \textit{correct} if it passes all checks from two widely used static analyzers, Bandit~\cite{bandit} and CodeQL~\cite{codeql}; otherwise, \textit{incorrect}.
\end{itemize}
Table~\ref{tab:dataset_com_func} and Table~\ref{tab:dataset_security} summarize the numbers of correct and incorrect code generated by each studied model for the given benchmarks and criteria.
The complete dataset, including generated code, the corresponding internal representations extracted from the studied code LLMs, and their correctness labels, could be found on our website~\cite{website}.

\begin{table}[!htp]\centering
\caption{The number of correct (\ncor) and incorrect (\ninc) solutions in \textit{Compilability} and \textit{Functionality} across benchmarks}\label{tab:dataset_com_func}
\resizebox{\columnwidth}{!}{ 
\begin{tabular}{l|l|l|r|r|rr}\toprule
Criteria & Code LLM & &\textit{HumanEval} &\textit{MBPP} &\textit{DevEval} \\\midrule
\multirow{12}{*}{Compilability} &\deepseek-1.3B &\ncor &1,464 &4,272 &5,952 \\
& &\ninc &176 &494 &11,377 \\\cmidrule{2-6}
&\deepseek-6.7B &\ncor &1,577 &4,635 &8,191 \\
& &\ninc &63 &147 &9,139 \\\cmidrule{2-6}
&\codellama-7B &\ncor &1,479 &3,804 &7,254 \\
& &\ninc &161 &824 &10,076 \\\cmidrule{2-6}
&\codellama-13B &\ncor &1,385 &4,128 &5,440 \\
& &\ninc &255 &792 &8,480 \\\cmidrule{2-6}
&\codegemma-7B &\ncor &1,315 &4,086 &5,049 \\
& &\ninc &325 &864 &8,870 \\\cmidrule{2-6}
&\magiccoder-7B &\ncor &1,572 &4,745 &7,432 \\
& &\ninc &66 &142 &9,898 \\\midrule

\multirow{12}{*}{Functionality} &\deepseek-1.3B &\ncor &872 &2,065 &2,092 \\
& &\ninc &768 &2,701 &15,237 \\\cmidrule{2-6}
&\deepseek-6.7B &\ncor &1,294 &3,033 &4,321 \\
& &\ninc &346 &1,749 &13,009 \\\cmidrule{2-6}
&\codellama-7B &\ncor &644 &1,942 &3,467 \\
& &\ninc &996 &2,686 &13,863 \\\cmidrule{2-6}
&\codellama-13B &\ncor &666 &2,065 &2,849 \\
& &\ninc &974 &2,855 &14,481 \\\cmidrule{2-6}
&\codegemma-7B &\ncor &718 &2,141 &1,876 \\
& &\ninc &922 &2,809 &12,043 \\\cmidrule{2-6}
&\magiccoder-7B &\ncor &1,198 &3,135 &4,113 \\
& &\ninc &440 &1,752 &13,217 \\
\bottomrule
\end{tabular}
}
\end{table}

\begin{table}[!htp]\centering
\caption{The number of correct (\ncor) and incorrect (\ninc) solutions in \textit{Security} across benchmarks}\label{tab:dataset_security}
\resizebox{\columnwidth}{!}{
\begin{tabular}{l|l|r|r|r|rr}\toprule
Code LLM & &\textit{SecurityEval} &\textit{CWEval} &\textit{CODEGUARD+} &\textit{SALLM} \\\midrule
\multirow{2}{*}{\deepseek-1.3B} &\ncor &482 &168 &200 &231 \\
&\ninc &728 &82 &510 &769 \\\midrule
\multirow{2}{*}{\deepseek-6.7B} &\ncor &508 &168 &267 &342 \\
&\ninc &702 &82 &443 &658 \\\midrule
\multirow{2}{*}{\codellama-7B} &\ncor &882 &221 &506 &750 \\
&\ninc &328 &29 &204 &250 \\\midrule
\multirow{2}{*}{\codellama-13B} &\ncor &675 &225 &519 &545 \\
&\ninc &535 &25 &191 &455 \\\midrule
\multirow{2}{*}{\codegemma-7B} &\ncor &597 &175 &289 &421 \\
&\ninc &613 &75 &421 &579 \\\midrule
\multirow{2}{*}{\magiccoder} &\ncor &598 &192 &262 &451 \\
&\ninc &612 &58 &448 &549 \\
\bottomrule
\end{tabular}
}
\end{table}
\subsection{Studied Code LLMs}

To enable a fair, reproducible, and practical evaluation, we selected the studied LLMs based on several key criteria. 
First, \textit{access to the model's  hidden states is essential}, as our goal is to evaluate code correctness by leveraging these internal representations. This requirement is typically met by open-source models, whereas closed-models or commercial models often restrict such access.
Second, \textit{we focus on code-specialized LLMs}, which are trained or fine-tuned on large-scale code corpora. These models generally outperform general-purpose LLMs on programming tasks, leading to a more realistic and meaningful evaluation.
Third, \textit{we consider the diversity of model architectures, model sizes, and training strategies} to broaden the scope of our analysis and improve the generalizability of our findings. 
Finally, \textit{we limit the model size to a maximum of 13 billion parameters} to accommodate hardware resource constraints and ensure that our evaluation is practical and reproducible.

Based on these criteria, we selected six representative LLMs from different model families to ensure both architectural diversity and broad generalizability of our evaluation, including \deepseek-1.3B~\cite{deepseek-coder}, \deepseek-6.7B~\cite{deepseek-coder}, \codellama-7B~\cite{codellama}, \codellama-13B~\cite{codellama},  \magiccoder-7B~\cite{magicoder}, and \codegemma-7B~\cite{codegemma}. 
All selected models are open-source, allowing full access to their internal representations during code generation. 
These models are widely adopted in both academic research and industrial applications, and have demonstrated strong performance on SE tasks~\cite{zheng2025towards, liu2023your, openia}.
Each employed model is either trained or fine-tuned on large-scale code corpora to capture the syntactic, semantic, and logical aspects of programming languages. 
For consistency, we use the instruction-tuned versions with officially released pre-trained weights from HuggingFace and do not perform further fine-tuning. This setup enables a fair assessment of each model's generalization ability and facilitates a controlled comparison of how their internal representations relate to code correctness.

\subsection{Baselines}
We compare the performance of \tool against SOTA \textit{white-box} and \textit{black-box} approaches. White-box approaches analyze the model's internal computations during code generation to determine the correctness of the generated output. In contrast, black-box approaches evaluate the code correctness based solely on the final generated output without considering the generation process.

\begin{itemize}
    \item \textit{White-box methods}~\cite{openia, inside, lookback}: These methods utilize internal signals such as attention maps, hidden activations, or internal representations as indicators of correctness. In this study, we evaluate the effectiveness of \tool by comparing it against a SOTA white-box method \openia~\cite{openia}, and an \oracle baseline.
    \begin{itemize}
        \item \openia~\cite{openia} determines if LLM-generated code is (in)correct by training a probing classifier using the internal states from the last generated token at the last layer of the models.

        \item \oracle represents an upper-bound baseline constructed by exhaustively searching across all layers and token positions to identify the \textit{most optimal} internal representations of each model for each evaluation setting. Since the search space grows exponentially with the number of layers and tokens, such a full search is computationally prohibitive and infeasible in practice. In this research, we approximate this oracle by individually evaluating representations from four boundary tokens (\texttt{first token}, \texttt{last token}, \texttt{first code token}, \texttt{last code token}) across layers, providing a  manageable yet informative estimation of the optimal choice.
    \end{itemize}

    \item \textit{Black-box methods}~\cite{openia, embedding_emse22,opt-pretrained-model,pretrained-survey, internal-state-2}: We consider two representative black-box strategies:
    \begin{itemize}
        
        \item \textit{Post-hoc classification-based methods}~\cite{embedding_emse22,opt-pretrained-model,pretrained-survey}: These methods predict correct code by considering the code semantics, typically captured using a pretrained model for code, e.g., CodeBERT~\cite{codebert}, CodeT5~\cite{codet5}, or CuBERT~\cite{cubert}. In this work, we employed two widely used models, including CodeBERT and CodeT5+, to encode the code semantics and train a classifier on the resulting embeddings to determine correctness.
        
        \item \textit{LLM-as-A-Judge (LLM-AJ)}~\cite{openia, internal-state-2}: These methods evaluate code correctness by utilizing the reasoning and generalization capabilities of LLMs. Following the prior work~\cite{openia, internal-state-2}, we adopt a zero-shot prompting strategy with two configurations: \inhouse, where the same model, that generates code, is used to assess its own output, and \external, where an external LLM (i.e., \gpt) serves as a verifier. The full prompt templates used in our experiments are available on our website~\cite{website}.
    \end{itemize}
\end{itemize}

\subsection{Evaluation Procedure}

\textbf{RQ1. Performance analysis:}
We evaluate \tool and the baselines across three key dimensions: \textit{compilability}, \textit{functionality}, and \textit{security} under different settings: 

\begin{itemize}
    \item \textbf{\textit{Independent-unit code generation}:} This focuses on standalone programming tasks~\cite{HumanEval, MBPP, SecurityEval, Cweval, CODEGUARD+, Sallm}.  For each studied LLM, the classifier is trained on internal representations extracted from one benchmark (e.g., MBPP) and evaluated on another (e.g., HumanEval). This setting tests the generalizability of correctness assessment across different benchmarks.

    \item \textbf{\textit{Repo(sitory)-level code generation}:} This setting evaluates the code correctness in the context of real-world software projects~\cite{li-etal-2024-deveval}. Repo-level tasks require LLMs to maintain contextual consistency across multiple code units, significantly increasing the complexity of code generation~\cite{rambo, zhang2023repocoder}. This setting measures how robust the approaches are in handling context-dependent code generation scenarios. 

\end{itemize}
For a fair comparison, we use the same architecture for the classification models in the post-hoc methods, \openia, and \tool. Each classifier contains an input layer, two hidden layers with 128 and 64 neurons respectively, and an output layer. All models are trained for 50 epochs using a batch size of 32 and a learning rate of $10^{-3}$.

\textbf{RQ2. Intrinsic Analysis:} We investigate how different components of \tool contribute to its overall performance. To answer this question, we systematically evaluate \tool under different configurations.

For the \textit{Internal Representation Extractor}, we aim to examine the impact of different sampling strategies:
\begin{itemize}
    \item \textit{Token sampling}: We compare \tool's performance under three sampling strategies: (1) \textit{full token}, (2) \textit{random}, and (3) \textit{boundary-aware}. \textit{Full token} uses representations from all generated tokens. To accommodate varying sequence lengths, we fix the sequence size to 256 tokens and apply padding or truncation as needed. \textit{Random} randomly selects one interior token position. \textit{Boundary-aware} selects four key tokens at boundary positions (Sec.~\ref{sec:extractor}).
  
    \item \textit{Layer sampling}: We evaluate the impacts of sampling different subsets of layers by varying the sampling intervals,  $k \in \{1, 2, 3, 4, 5\}$, which defines the step size between selected layers.
\end{itemize}

For the \textit{Informative Representation Selector}, we conduct an ablation study to investigate its contribution to \tool's performance. Specifically, we compare \tool's performance under two configurations:
\begin{itemize}
    \item \textit{With Selector}: This module is \textit{enabled} to assign importance scores to each internal representation.
    \item \textit{Without Selector}: The selector module is \textit{disabled}. All internal representations are uniformly aggregated with equal importance.
\end{itemize}

For the \textit{Code Correctness Predictor}, we study how the choice of aggregation function and probing classifier affects the overall results. Regarding the \textit{aggregation function}, we evaluate several strategies for combining the selected internal representations $\mathcal{Z}^*$, including concatenation, summation, mean pooling, and max pooling. For the \textit{probing classifier}, we explore various classifier architectures, including Multi-Layer Perceptron (MLP), Support Vector Machine (SVM), and Logistic Regression. 

All intrinsic analysis experiments are conducted using \codegemma-7B on the MBPP benchmark. For each experiment, the MBPP tasks are randomly split into a 90:10 ratio for training and testing. 
Internal representations extracted during the code generation process for the tasks in the training set are used to train the probing classifier, while those from the test set are used to evaluate its performance.

\textbf{RQ3. Sensitivity analysis:}
This experiment examines \tool's generalization capability across programming languages and LLMs. 
Specifically, we assess whether \tool, when trained on internal representations extracted from code generation in certain programming languages, can accurately predict the correctness of code written in a completely different language. 
We also analyze how varying the size of the training data affects \tool's prediction performance. 
Additionally, we examine \tool's ability to transfer across models by training it on the internal states of several code LLMs and testing its performance on code generated by a different model.

\subsection{Metrics}

Determining whether code generated by an LLM is correct can be formulated as a binary classification problem. 
To evaluate \tool and the baseline methods, we employ standard classification metrics. 
For a fair evaluation across both classes, especially in the presence of class imbalance, we adopt weighted metrics including \textit{Accuracy}, \textit{Weighted Precision}, \textit{Weighted Recall}, and \textit{Weighted F1-Score}.
These metrics compute a label-wise average weighted by the proportion of each class in the test set.
We also evaluate the efficiency of each approach by measuring time complexity (for both training and inference) and memory usage. These metrics provide a comprehensive understanding of the practical feasibility of the approaches.
\section{Experimental Results}

\subsection{Performance Analysis}

\subsubsection{Security Assessment}

\begin{table*}\centering
\caption{{Correctness assessment performance in \textit{\textbf{security}} criterion }}
\label{tab:rq1_performance_security}
\begin{tabular}{l|l|rrrrr}\toprule
Code LLM & &Accuracy &Precision &Recall &F1-Score \\\midrule
\multirow{7}{*}{\deepseek-1.3B}

&\cellcolor[HTML]{ffe599}\oracle &\cellcolor[HTML]{ffe599}0.88 &\cellcolor[HTML]{ffe599}0.88 &\cellcolor[HTML]{ffe599}0.88 &\cellcolor[HTML]{ffe599}0.88  \\

&\inhouse &0.38 &0.15 &0.38 &0.21 \\
&\external &0.63 &0.63 &0.63 &0.63 \\
&CodeBERT &0.81 &0.81 &0.81 &0.81 \\
&CodeT5+ &0.80 &0.80 &0.80 &0.79 \\
&\openia &0.81 &0.81 &0.81 &0.81 \\
&\tool &\textbf{0.90} &\textbf{0.90} &\textbf{0.90} &\textbf{0.89} \\

\midrule
\multirow{7}{*}{\deepseek-6.7B } 

&\cellcolor[HTML]{ffe599}\oracle &\cellcolor[HTML]{ffe599}0.84 &\cellcolor[HTML]{ffe599}0.84 &\cellcolor[HTML]{ffe599}0.84 &\cellcolor[HTML]{ffe599}0.84  \\

&\inhouse &0.48 &0.23 &0.48 &0.32 \\
&\external &0.69 &0.70 &0.69 &0.69 \\
&CodeBERT &0.80 &0.80 &0.80 &0.80 \\
&CodeT5+ &0.80 &0.81 &0.80 &0.79 \\
&\openia &0.81 &0.81 &0.81 &0.81 \\
&\tool &\textbf{0.82} &\textbf{0.82} &\textbf{0.82} &\textbf{0.82} \\

\midrule
\multirow{7}{*}{\codellama-7B} 
&\cellcolor[HTML]{ffe599}\oracle &\cellcolor[HTML]{ffe599}0.94 &\cellcolor[HTML]{ffe599}0.95 &\cellcolor[HTML]{ffe599}0.94 &\cellcolor[HTML]{ffe599}0.94  \\
&\inhouse &0.73 &0.53 &0.73 &0.62 \\
&\external &0.50 &0.70 &0.50 &0.51 \\
&CodeBERT &0.85 &0.89 &0.85 &0.86 \\
&CodeT5+ &0.86 &0.88 &0.86 &0.86 \\
&\openia &0.87 &0.87 &0.87 &0.87 \\
&\tool &\textbf{0.91} &\textbf{0.92} &\textbf{0.91} &\textbf{0.91} \\

\midrule
\multirow{7}{*}{\codellama-13B} 
&\cellcolor[HTML]{ffe599}\oracle &\cellcolor[HTML]{ffe599}0.89 &\cellcolor[HTML]{ffe599}0.89 &\cellcolor[HTML]{ffe599}0.89 &\cellcolor[HTML]{ffe599}0.89  \\
&\inhouse &0.64 &0.40 &0.64 &0.49 \\
&\external &0.54 &0.65 &0.54 &0.53 \\
&CodeBERT &0.82 &0.84 &0.82 &0.83 \\
&CodeT5+ &0.83 &0.84 &0.83 &0.83 \\
&\openia &0.75 &0.75 &0.75 &0.75 \\
&\tool &\textbf{0.88} &\textbf{0.88} &\textbf{0.88} &\textbf{0.88} \\

\midrule
\multirow{7}{*}{\magiccoder-7B } 
&\cellcolor[HTML]{ffe599}\oracle &\cellcolor[HTML]{ffe599}0.87 &\cellcolor[HTML]{ffe599}0.87 &\cellcolor[HTML]{ffe599}0.87 &\cellcolor[HTML]{ffe599}0.87  \\
&\inhouse &0.51 &0.26 &0.51 &0.35 \\
&\external &0.65 &0.66 &0.65 &0.64 \\
&CodeBERT &0.75 &0.76 &0.75 &0.75 \\
&CodeT5+ &0.79 &0.79 &0.79 &0.79 \\
&\openia &0.73 &0.73 &0.73 &0.73 \\
&\tool &\textbf{0.85} &\textbf{0.85} &\textbf{0.85} &\textbf{0.85} \\

\midrule
\multirow{7}{*}{\codegemma-7B} 
&\cellcolor[HTML]{ffe599}\oracle &\cellcolor[HTML]{ffe599}0.89 &\cellcolor[HTML]{ffe599}0.89 &\cellcolor[HTML]{ffe599}0.89 &\cellcolor[HTML]{ffe599}0.89  \\
&\inhouse &0.51 &0.26 &0.51 &0.35 \\
&\external &0.63 &0.65 &0.63 &0.62 \\
&CodeBERT &0.78 &0.78 &0.78 &0.78 \\
&CodeT5+ &0.77 &0.78 &0.77 &0.77 \\
&\openia &0.77 &0.79 &0.77 &0.77 \\
&\tool &\textbf{0.88} &\textbf{0.88} &\textbf{0.88} &\textbf{0.88} \\

\bottomrule
\end{tabular}
\end{table*}

The performance of \tool and the baseline approaches in code correctness assessment under the security criterion is shown in Table~\ref{tab:rq1_performance_security}. Overall, \textit{\tool consistently outperforms all baselines across all studied code LLMs. Notably, \tool's performance closely matches or even exceeds that of \oracle, indicating its effectiveness in capturing informative internal signals.}

Compared to the white-box approach \openia, \tool yields improvement gains of up to 18\%. For example, when assessing code generated by \codellama-13B, \tool reaches an accuracy of 0.88, while this figure of \openia is only 0.75. This performance gap can be attributed to the limitations of \openia's design, which relies on fixed representations from the last generated token at the last layer of the LLM. Such a rigid strategy hinders its flexibility and generalizability across different model architectures.
In contrast, \tool dynamically selects the most informative internal states,  allowing it to tailor the selected representations to each model's characteristics. This flexible mechanism enables \tool to capture correctness-related signals more effectively, resulting in more robust performance across diverse models.

Both \inhouse and \external achieve considerably low performance. The accuracy of \inhouse lags behind \tool by margins ranging from 24\% to 133\%.
In the \inhouse setting, the same LLM that generates the code is also used to assess its own outputs, making it prone to overconfidence in the correctness of its responses. Similarly, \external falls short of \tool by 19\% to 82\%, even when using a powerful external model like \gpt.
This performance gap underscores the challenge of detecting security vulnerabilities, which often require deep semantic understanding and context-aware reasoning.
These results expose the limitations of using LLMs directly as correctness judges under the security criterion. 

The back-box classification approaches based on CodeBERT and CodeT5+ show quite strong performance across all code LLMs in evaluating the security correctness. For instance, the accuracies achieved by CodeT5+ are comparable to \openia and better than \external by up to 55\%. However, their performance still lags behind \tool from 5\% to 14\%.

Furthermore, \tool exhibits strong performance, which is close to \oracle. 
Interestingly, on \deepseek-1.3B, \tool achieves an accuracy of 0.90, which slightly surpasses \oracle's performance. By aggregating the sampled and weighted internal representations, \tool is able to not only focus on the most informative states but also integrate complementary information across multiple tokens and layers. This mechanism enhances its ability to detect security vulnerabilities. As a result, \tool can outperform \oracle, which relies on a single representation in some cases.
These results highlight the effectiveness of \tool in capturing important signals for security correctness prediction.

\subsubsection{Functionality Assessment}

\begin{table*}\centering
\caption{{Correctness assessment performance in \textit{\textbf{functionality}} criterion }}
\label{tab:rq1_performance_functionality}
\begin{tabular}{l|l|rr|rrr}\toprule
Code LLM& &\multicolumn{2}{c}{Independent-unit code} &\multicolumn{2}{|c}{Repo-level code} \\\cmidrule{3-6}
 & &Accuracy &F1-Score &Accuracy &F1-Score \\
 \midrule
\multirow{7}{*}{\deepseek-1.3B} 
&\cellcolor[HTML]{ffe599}\oracle &\cellcolor[HTML]{ffe599}0.66 &\cellcolor[HTML]{ffe599}0.66 &\cellcolor[HTML]{ffe599}0.88 &\cellcolor[HTML]{ffe599}0.85 \\
&\inhouse &0.55 &0.50 &0.24 &0.27 \\
&\external &\textbf{0.93} &\textbf{0.93} &\textbf{0.87} &\textbf{0.87}\\
&CodeBERT &0.50 &0.45 &0.86 &0.81 \\
&CodeT5+ &0.47 &0.34 &0.86 &0.83 \\
&\openia &0.66 &0.66 &0.86 &0.83 \\
&\tool &0.63 &0.63 &\textbf{0.87} &0.84 \\

\midrule
\multirow{7}{*}{\deepseek-6.7B} 
&\cellcolor[HTML]{ffe599}\oracle &\cellcolor[HTML]{ffe599}0.77 &\cellcolor[HTML]{ffe599}0.72 &\cellcolor[HTML]{ffe599}0.77 &\cellcolor[HTML]{ffe599}0.75 \\
&\inhouse &0.32 &0.26 &0.48 &0.50 \\
&\external &\textbf{0.94} &\textbf{0.94} &0.70 &0.71 \\
&CodeBERT &0.35 &0.35 &0.71 &0.68 \\
&CodeT5+ &0.66 &0.66 &0.73 &0.69 \\
&\openia &0.65 &0.67 &0.75 &\textbf{0.73} \\
&\tool &0.71 &0.71 &\textbf{0.76} &\textbf{0.73} \\

\midrule
\multirow{7}{*}{\codellama-7B} 
&\cellcolor[HTML]{ffe599}\oracle &\cellcolor[HTML]{ffe599}0.69 &\cellcolor[HTML]{ffe599}0.69 &\cellcolor[HTML]{ffe599}0.81 &\cellcolor[HTML]{ffe599}0.79 \\
&\inhouse &0.61 &0.46 &0.51 &0.54 \\
&\external &\textbf{0.97} &\textbf{0.97} &0.73 &0.74 \\
&CodeBERT &0.58 &0.54 &0.74 &0.69 \\
&CodeT5+ &0.63 &0.62 &0.76 &0.71 \\
&\openia &0.69 &0.68 &\textbf{0.81} &\textbf{0.79} \\
&\tool &0.68 &0.68 &\textbf{0.81} &0.78 \\

\midrule
\multirow{7}{*}{\codellama-13B} 
&\cellcolor[HTML]{ffe599}\oracle &\cellcolor[HTML]{ffe599}0.74 &\cellcolor[HTML]{ffe599}0.73 &\cellcolor[HTML]{ffe599}0.82 &\cellcolor[HTML]{ffe599}0.79 \\
&\inhouse &0.41 &0.23 &0.19 &0.06 \\
&\external &\textbf{0.95} &\textbf{0.95} &0.77 &0.78 \\
&CodeBERT &0.61 &0.61 &0.80 &0.74 \\
&CodeT5+ &0.62 &0.62 &0.78 &0.73 \\
&\openia &0.64 &0.64 &\textbf{0.81} &0.78 \\
&\tool &0.68 &0.66 &\textbf{0.81} &\textbf{0.79} \\

\midrule
\multirow{7}{*}{\magiccoder-7B} 
&\cellcolor[HTML]{ffe599}\oracle &\cellcolor[HTML]{ffe599}0.74 &\cellcolor[HTML]{ffe599}0.69 &\cellcolor[HTML]{ffe599}0.78 &\cellcolor[HTML]{ffe599}0.74 \\
&\inhouse &0.27 &0.11 &0.74 &0.65 \\
&\external &\textbf{0.94} &\textbf{0.94} &0.70 &0.72 \\
&CodeBERT &0.46 &0.49 &0.73 &0.69 \\
&CodeT5+ &0.54 &0.56 &0.72 &0.69 \\
&\openia &0.56 &0.58 &0.77 &0.73 \\
&\tool &0.69 &0.66 &\textbf{0.81} &\textbf{0.78} \\

\midrule
\multirow{7}{*}{\codegemma-7B} 
&\cellcolor[HTML]{ffe599}\oracle &\cellcolor[HTML]{ffe599}0.70 &\cellcolor[HTML]{ffe599}0.70 &\cellcolor[HTML]{ffe599}0.84 &\cellcolor[HTML]{ffe599}0.76 \\
&\inhouse &0.39 &0.22 &0.16 &0.05 \\
&\external &\textbf{0.91} &\textbf{0.91} &0.79 &\textbf{0.80}\\
&CodeBERT &0.56 &0.53 &0.82 &0.79 \\
&CodeT5+ &0.59 &0.57 &0.79 &0.77 \\
&\openia &0.61 &0.59 &0.82 &0.77 \\
&\tool &0.65 &0.64 &\textbf{0.84} &0.77 \\

\bottomrule
\end{tabular}
\end{table*}

Table~\ref{tab:rq1_performance_functionality} shows that \textit{\external is the most effective approach for assessing the functionality correctness of \textit{independent-unit} code, while \tool performs best in the \textit{repo-level} code setting.} Specifically, in the \textit{independent-unit} setting, \external achieves an average accuracy of 0.94, exceeding that of the other approaches by 32--121\%.
In the \textit{repo-level} setting, \tool reaches an average accuracy of 0.82, which is comparable to the \oracle's performance and surpasses the other approaches by up to 111\%.


The performance of LLM-AJ approaches, including both \inhouse and \external, \textit{declines} when transitioning from \textit{independent-unit} setting to \textit{repo-level} setting. For example, when evaluating the functional correctness of standalone programs generated by \codegemma-7B, \inhouse and \external achieve F1-Scores of 0.22 and 0.91, respectively. However, these scores significantly drop to 0.05 and 0.80 when assessing the correctness of \textit{repo-level} code.
This performance degradation can be attributed to the \textit{reasoning-based} nature of LLM-AJ approaches. Specifically, they judge functional correctness by analyzing and reasoning over the provided context. In the \textit{independent-unit} setting, the program is self-contained, allowing LLMs to access to the full context needed for reliable reasoning. Meanwhile, \textit{repo-level} code involves multi-file dependencies, preventing LLMs from explicitly accessing the entire project. As a result, they lack crucial information, leading to degraded performance in the \textit{repo-level} setting.

In contrast, the other approaches shows performance \textit{gains} when moving from \textit{independent-unit} to \textit{repo-level} functionality assessment. For instance, the black-box method with CodeBERT improves by 52\%, while \tool achieves a 21\% increase. 
These methods,
including black-box classifiers, \openia, and \tool, are \textit{representation-based} approaches that distinguish correct and incorrect code by leveraging features encoded in embedding vectors or hidden states of the generated code. In the \textit{independent-unit} setting, generated code snippets are often short and simple, offering limited  discriminative signals. This makes it more challenging for the representation-based models to separate correct and incorrect cases. On the other hand, the \textit{repo-level} setting involves longer and more complex code, which introduces richer structural and semantic information. 
These characteristics lead to more distinctive feature patterns in the representations, thereby improving separability. 
As a result, representation-based approaches demonstrate improved performance of functionality assessment when transitioning from the \textit{independent-unit} to the \textit{repo-level} setting.

Overall, these differences highlight that reasoning-based methods are highly dependent on complete contextual information, whereas representation-based approaches can exploit the richer feature patterns of complex code to achieve better robustness in \textit{repo-level} functionality assessment.


\begin{figure*}
    \centering
\includegraphics[width=\linewidth]{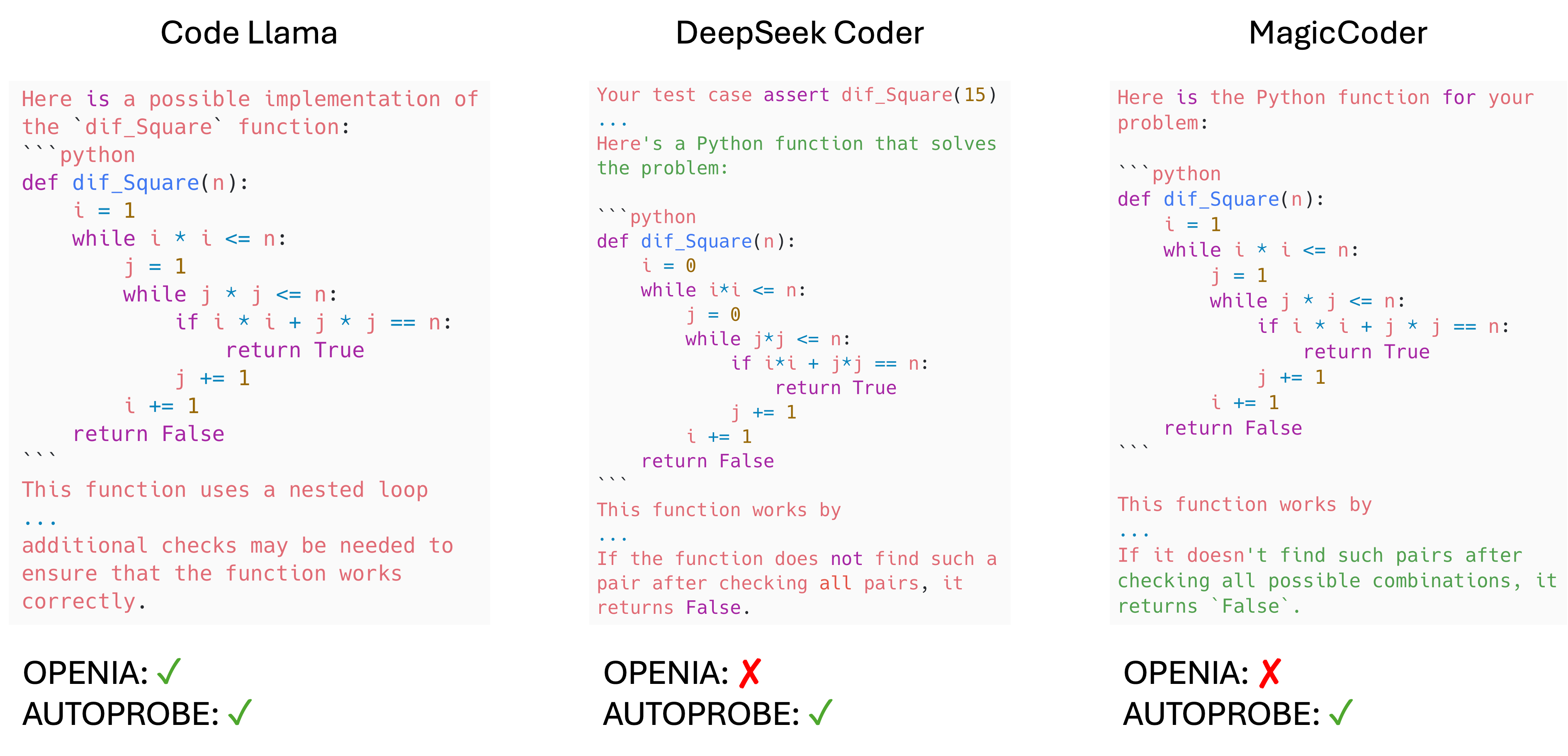}
    \caption{An example of code generated by \codellama-7B, \deepseek-6.7B, and \magiccoder-7B for Task ID\_72 in the MBPP benchmark. The task requires ``\texttt{Write a Python function to check whether the given number can be represented as difference of two squares or not}''. Due to space constraints, portions of the generated outputs are omitted. Full generations are available on our website~\cite{website}.
    }
    \label{fig:example}
\end{figure*}

Furthermore, compared to the SOTA white-box approach \openia, \tool exhibits greater robustness. 
Figure~\ref{fig:example} shows an example of code generated by \codellama-7B, \deepseek-6.7B, and \magiccoder-7B for the same task.
Although all the three models produced similar implementations, the generated functions are all incorrect. Specifically, instead of checking whether a number can be represented as the \textit{difference} of two squares, the generated code mistakenly checks for the \textit{sum} of two squares.

Despite the nearly identical functions, \openia yields inconsistent predictions about their functional correctness. It accurately assesses the correctness of the code generated by \codellama-7B, but  misclassified the outputs from \deepseek-6.7B and \magiccoder-7B. This inconsistency arises because \openia relies on the \textit{fixed} representation of only the last generated token (e.g., token \texttt{correctly} for \codellama and \texttt{False} for the other models) at the final layer of the models. Since different model architectures and generation dynamics assign varying semantic signals to these tokens, \openia's predictions become unstable and unreliable across models, even when the generated code is essentially the same.

Meanwhile, \tool overcomes this limitation through an adaptive mechanism that dynamically selects the most informative internal signals for each model. This flexibility allows \tool to remain robust to architectural differences and consistently produce correct predictions for the code generated by all three models. As a result, \tool achieves reliable correctness assessment across diverse code LLMs, demonstrating its superiority in both stability and effectiveness over \openia.
%

\subsubsection{Compilability Assessment}

\begin{table*}\centering
\caption{{Correctness assessment performance in \textit{\textbf{compilability}} criterion }}
\label{tab:rq1_performance_compilability}
\begin{tabular}{l|l|rr|rrr}\toprule
Code LLM& &\multicolumn{2}{c}{Independent-unit code} &\multicolumn{2}{|c}{Repo-level code} \\\cmidrule{3-6}
 & &Accuracy &F1-Score &Accuracy &F1-Score \\
 \midrule
\multirow{7}{*}{\deepseek-1.3B} 

&\cellcolor[HTML]{ffe599}\oracle &\cellcolor[HTML]{ffe599}0.90 &\cellcolor[HTML]{ffe599}0.86 &\cellcolor[HTML]{ffe599}0.67 &\cellcolor[HTML]{ffe599}0.67  \\
&\inhouse &0.53 &0.37 &0.48 &0.32 \\
&\external &\textbf{0.91} &\textbf{0.89} &0.55 &0.50 \\
&CodeBERT &0.84 &0.83 &0.61 &0.60 \\
&CodeT5+ &0.88 &0.84 &0.61 &0.59 \\
&\openia &0.88 &0.85 &0.60 &0.59 \\
&\tool &\text{0.89} &\text{0.85} &\textbf{0.65} &\textbf{0.65} \\

\midrule
\multirow{7}{*}{\deepseek-6.7B} 
&\cellcolor[HTML]{ffe599}\oracle &\cellcolor[HTML]{ffe599}0.96 &\cellcolor[HTML]{ffe599}0.94 &\cellcolor[HTML]{ffe599}0.68 &\cellcolor[HTML]{ffe599}0.68  \\

&\inhouse &0.76 &0.65 &0.61 &0.46 \\
&\external &0.94 &0.94 &0.61 &0.55 \\
&CodeBERT &0.94 &0.93 &0.57 &0.58 \\
&CodeT5+ &\textbf{0.96} &0.94 &0.58 &0.59 \\
&\openia &\textbf{0.96} &0.94 &0.61 &0.61 \\
&\tool &\textbf{0.96} &\textbf{0.95} &\textbf{0.64} &\textbf{0.64} \\

\midrule
\multirow{7}{*}{\codellama-7B} 
&\cellcolor[HTML]{ffe599}\oracle &\cellcolor[HTML]{ffe599}0.90 &\cellcolor[HTML]{ffe599}0.86 &\cellcolor[HTML]{ffe599}0.66 &\cellcolor[HTML]{ffe599}0.66  \\

&\inhouse &0.39 &0.22 &0.58 &0.42 \\
&\external &\textbf{0.91} &\textbf{0.88} &0.60 &0.55 \\
&CodeBERT &0.87 &0.84 &0.55 &0.55 \\
&CodeT5+ &0.56 &0.65 &0.60 &0.59 \\
&\openia &0.88 &0.86 &0.60 &0.61 \\
&\tool &0.90 &0.86 &\textbf{0.66} &\textbf{0.66} \\

\midrule
\multirow{7}{*}{\codellama-13B} 
&\cellcolor[HTML]{ffe599}\oracle &\cellcolor[HTML]{ffe599}0.90 &\cellcolor[HTML]{ffe599}0.88 &\cellcolor[HTML]{ffe599}0.70 &\cellcolor[HTML]{ffe599}0.71  \\

&\inhouse &0.41 &0.23 &0.57 &0.42 \\
&\external &\textbf{0.91} &\textbf{0.91} &0.60 &0.54 \\
&CodeBERT &0.81 &0.79 &0.60 &0.59 \\
&CodeT5+ &0.90 &0.88 &0.62 &0.62 \\
&\openia &0.87 &0.87 &0.66 &0.66 \\
&\tool &0.90 &0.87 &\textbf{0.69} &\textbf{0.69} \\

\midrule
\multirow{7}{*}{\magiccoder-7B} 
&\cellcolor[HTML]{ffe599}\oracle &\cellcolor[HTML]{ffe599}0.96 &\cellcolor[HTML]{ffe599}0.94 &\cellcolor[HTML]{ffe599}0.68 &\cellcolor[HTML]{ffe599}0.69  \\

&\inhouse &0.73 &0.62 &0.60 &0.45 \\
&\external &0.95 &\textbf{0.95} &\textbf{0.66} &0.63 \\
&CodeBERT &0.94 &0.93 &0.58 &0.58 \\
&CodeT5+ &0.88 &0.90 &0.57 &0.57 \\
&\openia &\textbf{0.96} &0.94 &0.61 &0.61 \\
&\tool &\textbf{0.96} &0.94 &\textbf{0.66} &\textbf{0.67} \\

\midrule
\multirow{7}{*}{\codegemma-7B} 
&\cellcolor[HTML]{ffe599}\oracle &\cellcolor[HTML]{ffe599}0.88 &\cellcolor[HTML]{ffe599}0.86 &\cellcolor[HTML]{ffe599}0.68 &\cellcolor[HTML]{ffe599}0.68  \\

&\inhouse &0.44 &0.27 &0.52 &0.36 \\
&\external &0.88 &0.86 &0.55 &0.49 \\
&CodeBERT &0.81 &0.82 &0.58 &0.56 \\
&CodeT5+ &0.78 &0.80 &0.61 &0.60 \\
&\openia &0.84 &0.79 &0.67 &0.67 \\
&\tool &\textbf{0.90} &\textbf{0.89} &\textbf{0.69} &\textbf{0.69} \\
\bottomrule
\end{tabular}
\end{table*}

Table~\ref{tab:rq1_performance_compilability} shows the results of the compilability assessment under two levels of code generation granularity, \textit{independent-unit} code and \textit{repo-level} code. As seen, \textit{most correctness assessment approaches achieve strong performance in \textit{independent-unit} code, yet show considerable decline in the more complex \textit{repo-level} setting}. Among them, \textit{\tool demonstrates the highest robustness, maintaining the smallest performance drop between the two settings.}

Specifically, in the \textit{independent-unit} setting, \external achieves the highest performance, with an average accuracy of 0.92, which is 69\% higher than \inhouse and 11\% higher than black-box classification with CodeT5+. However, its performance drops by 1.5 times, reaching an average accuracy of 0.59 in the \textit{repo-level} setting. 
Indeed, determining whether a piece of code is compilable is mainly dependent on checking syntax and structural correctness, without the need for deep reasoning. 
In \textit{independent-unit} snippets, all relevant information, such as imports, declared variables, and function definitions, is self-contained, allowing LLMs like \gpt to simply scan the code to verify its syntactic validity. 
Meanwhile, \textit{repo-level} code often references variables, functions, or modules defined across multiple files. Without explicit access to the entire project context, the LLM struggles to accurately determine compilability, leading to a significant performance drop in this setting.

In the more complex setting of \textit{repo-level} code assessment, \tool achieves the highest performance, outperforming \external by 12\% on average. 
For example, when evaluating the compilability of code generated by \codegemma-7B, both \tool and \external reach the accuracy of 0.90 on \textit{independent-unit} code. 
However, for \textit{repo-level} code, \tool obtains an accuracy of 0.69, which is 27\% higher than that of \external. Interestingly, in this case, \tool even slightly surpasses the performance of \oracle.
These results demonstrate that \tool is less sensitive to the increased code complexity, making it a more reliable choice for practical correctness assessment scenarios.

\begin{gtheorem} 
\textbf{Answer to RQ1}: Across all three correctness dimensions, security, functionality, and compilability, \tool demonstrates strong and consistent performance, outperforming  SOTA baselines.
\end{gtheorem}
\subsection{Intrinsic Analysis}

\subsubsection{Impact of Representation Sampling Strategies}
\label{sec:impact_sampling}
\begin{figure}
    \centering
    \includegraphics[width=\columnwidth]{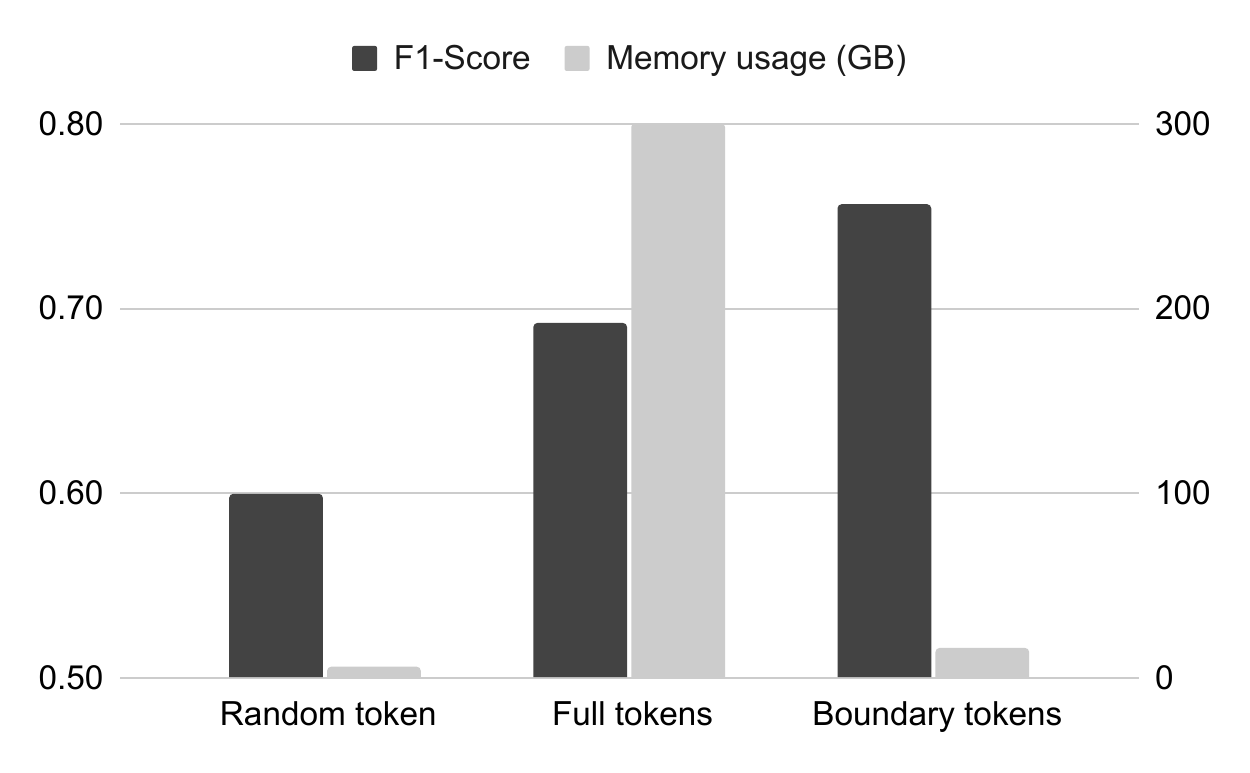}
    \caption{Impact of token sampling on \tool's performance and memory usage, left axis: \textit{F1-Score}; right axis: \textit{Memory usage (GB)}}
    \label{fig:token_sampling}
\end{figure}

\begin{figure}
    \centering
    \includegraphics[width=\columnwidth]{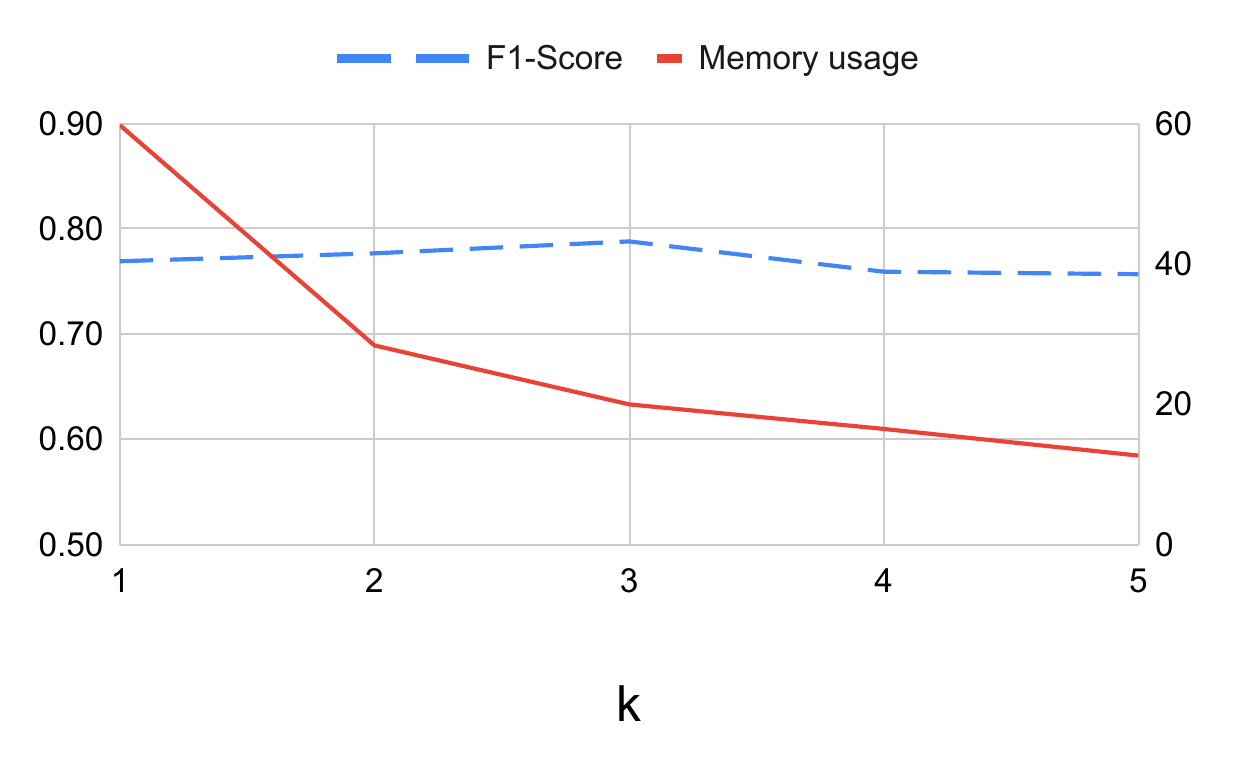}
    \caption{Impact of layer sampling intervals $k$ on \tool's performance and memory usage, left axis: \textit{F1-Score}; right axis: \textit{Memory usage (GB)}}
    \label{fig:layer_sampling}
\end{figure}

\textbf{Token sampling:}
Figure~\ref{fig:token_sampling} shows the impact of different token sampling strategies on \tool's performance. Overall, \textit{the boundary-aware strategy offers the best balance, achieving the highest F1-Score with relatively low memory consumption.}
While random token sampling results in the lowest memory usage (about 6.5 GB across GPU and CPU), it yields the weakest performance. This is expected because randomly selecting a single token may overlook important signals, leading to suboptimal predictions.

In contrast, \tool with boundary-aware token sampling consumes 2.5 times more memory than random sampling, but achieves significantly better performance. Specifically, \tool reaches an F1-Score of 0.76, outperforming full token and random token sampling strategies by 9\% and 26\%, respectively. Moreover, compared to full token sampling, the boundary-aware strategy reduces memory usage by 18X and training time by over 100X.
These results suggest that using all tokens introduces substantial computational overhead and even degrades performance due to the inclusion of noisy and irrelevant representations. Meanwhile, the boundary-aware strategy effectively reduces input size while preserving the most informative signals for correctness assessment.

\textbf{Layer Sampling:}
Figure~\ref{fig:layer_sampling} shows how the layer sampling interval $k \in \{1, 2, 3, 4, 5\}$, i.e., how frequently layers are selected for extracting internal states, affects \tool's performance and memory usage.
\textit{Across different sampling intervals, \tool's F1-Scores are relatively stable}, fluctuating between 0.76 and 0.79.  
This stability can be explained by the fact that adjacent layers in transformer-based models tend to encode highly similar features,  with similarity increasing as layers get closer~\cite{jiang2024tracing}. As a result, densely sampling layers (small $k$) often captures redundant features without providing additional information for correctness prediction.

Meanwhile, \textit{memory usage decreases steadily as $k$ increases,} since larger intervals result in fewer selected layers and a smaller set of extracted representations $\mathcal{H}^*$, thereby lowering memory consumption.
For example, when $k = 1$, \tool requires about 60 GB of GPU and CPU, which is 4.7X more than when $k=5$. 
These results highlight the importance of selecting a representative subset of layers rather than using all of them, which increases overhead without improving code correctness prediction performance.
\subsubsection{Impact of Informative Representation Selector}

\begin{table}[]\centering
\caption{Impact of \textit{Informative Representation Selector} on \tool's Performance}
\label{tab:intrinsic_selector}
\resizebox{\columnwidth}{!}{ 
\begin{tabular}{l|rrrrr}\toprule
 &Accuracy &Precision &Recall &F1-Score \\
\midrule
Without selector &0.70 &0.70 &0.70 &0.69 \\
With selector  &0.76 &0.76 &0.76 &0.76 \\
\bottomrule
\end{tabular}
}
\end{table}
Table~\ref{tab:intrinsic_selector} illustrates that \textit{the informative representation selector module significantly contributes to \tool's performance}. 
Indeed, not all internal representations contribute equally to correctness prediction. Without a mechanism to identify and prioritize useful signals, the classifier struggles to capture important information.
By incorporating an attention mechanism to assign higher weights to the important representations and downweight the irrelevant ones, \tool's F1-Score in code correctness prediction improves from 0.69 to 0.76 (more than 10\% relatively). 
The selector module thus plays a crucial role in directing the model's focus toward the most relevant internal states, leading to more effective and reliable correctness assessment. 
Moreover, since the attention scores are jointly trained with the probing classifier, the training time remains comparable with and without the selector module.
This suggests that the performance improvement comes at a reasonable computational cost, making the selector a lightweight yet useful addition to the \tool's design.

\subsubsection{Impact of Aggregation and Probing Classier}
\label{sec:impact_aggregation_classifier}

\begin{table}[!htp]\centering
\caption{Impacts of Aggregation Functions on \tool's Performance}
\label{tab:intrinsic_aggregation}
\resizebox{\columnwidth}{!}{ 
\begin{tabular}{l|rrrrr}\toprule
Aggregation &Accuracy &Precision &Recall &F1-Score \\\midrule
Concatenation &0.76 &0.76 &0.76 &0.76 \\
Summation &0.76 &0.76 &0.76 &0.76 \\
Mean pooling &0.76 &0.76 &0.76 &0.76 \\
Max pooling &0.79 &0.80 &0.79 &0.79 \\
Min pooling &0.73 &0.73 &0.73 &0.73 \\
\bottomrule
\end{tabular}
}
\end{table}

Table~\ref{tab:intrinsic_aggregation} presents the performance of \tool with different \textit{\textbf{aggregation functions}}, which are applied to combine the weighted representations in $\mathcal{Z}^*$ before feeding them to a probing classifier (Sec.~\ref{sec:predictor}). 
Among the evaluated strategies, \textit{max pooling proves to be the most effective aggregation function, enabling \tool to achieve the highest prediction performance.}
As seen, with \textit{max pooling}, \tool achieves an accuracy of 0.79, which is about 10\% higher than the worst setting using \textit{min pooling}. 
This indicates that \textit{max pooling}, by selecting the most prominent features across representation dimensions, captures more discriminative signals for code correctness prediction, thereby enhancing \tool's overall performance.

\begin{table}[!htp]\centering
\caption{Impact of Probing Classifiers on \tool's Performance}
\label{tab:intrinsic_classifiers}
\resizebox{\columnwidth}{!}{ 
\begin{tabular}{l|rrrrr}\toprule
\textbf{Classifier} &Accuracy &Precision &Recall &F1-Score \\\midrule
Logistic Regression &0.77 &0.77 &0.77 &0.77 \\
MLP &0.76 &0.76 &0.76 &0.76 \\
SVM &0.47 &0.22 &0.47 &0.30 \\
\bottomrule
\end{tabular}
}
\end{table}

The performance of different \textit{\textbf{probing classifiers}} used in \tool is shown in Table~\ref{tab:intrinsic_classifiers}. \textit{Both Logistic Regression and MLP achieve comparable results, while the SVM classifier performs significantly worse}. Specifically, \tool obtains an average accuracy of about 0.77 with either Logistic Regression or MLP, whereas this figure drops to only 0.47 when using SVM.
SVM is not a reliable choice in this context due to its limited capability in handling high-dimensional and possibly noisy feature spaces, which is an inherent characteristic of the internal representations extracted from LLMs. 
Furthermore, SVM is not typically suitable for joint training with upstream components such as the attention module, as their standard optimization is not gradient-based.
Although SVM can be trained using gradient-based methods via hinge loss, the loss function itself is non-smooth, which can lead to unstable or inefficient gradient propagation.
In contrast, Logistic Regression and MLP are trained using gradient descent and are fully differentiable, making them naturally compatible with \tool's end-to-end learning framework.

\begin{gtheorem} 
\textbf{Answer to RQ2}: Each component of \tool effectively contributes to its overall performance and robustness. Within the internal representation extractor, compact and structured subsets of internal states, which are obtained through suitable token and layer sampling strategies, significantly improve efficiency without sacrificing predictive accuracy. 
The informative representation selector and the max pooling aggregation function further enhance performance by capturing the most salient features from the selected representations. 
Finally, gradient-based classifiers such as Logistic Regression and MLP are well-suited for integration into \tool's joint training framework.
\end{gtheorem} 
\subsection{Sensitivity Analysis}

\subsubsection{Impact of Programming Language Variability}

\begin{table}[!htp]\centering
\caption{Correctness assessment performance in F1-score of the approaches across different programming languages}
\label{tab:sensitivity_languages}
\resizebox{\columnwidth}{!}{ 
\begin{tabular}{l|rrrrr}\toprule
Targeted language &CodeBERT &CodeT5+ &\openia &\tool \\\midrule
CPP &0.65 &0.67 &0.82 &0.76 \\
C Sharp &0.74 &0.75 &0.80 &0.85 \\
Java &0.59 &0.64 &0.68 &0.76 \\
JavaScript &0.66 &0.80 &0.86 &0.89 \\
PHP &0.54 &0.65 &\cellcolor[HTML]{f8f9fa}0.71 &0.83 \\
Python &0.37 &0.50 &0.67 &0.78 \\
Shell Script &0.58 &0.58 &0.64 &0.67 \\
TypeScript &0.72 &0.82 &0.88 &0.86 \\
\midrule
Average &0.61 &0.68 &0.76 &0.80 \\
\bottomrule
\end{tabular}
}
\end{table}

Table~\ref{tab:sensitivity_languages} shows the generalization performance of the approaches in assessing the code correctness across different programming languages. This experiment is conducted on the multilingual version of \textit{HumanEval} benchmark.
For each task in a target language, \deepseek-6.7B is employed to generate 10 candidate solutions. 
To evaluate cross-language generalization, one programming language is held out for testing, while the classifiers are trained on either the generated code (for post-hoc classifications) or internal representations (for \openia and \tool) from the remaining languages. 

Overall, \textit{\tool exhibits strong generalization capability across diverse programming languages}. \tool achieves the highest average F1-Score of 0.80, outperforming \openia by 5\% and the black-box methods by up to 32\%. 
Notably, for Python programs, \tool obtains an F1-score of 0.78, 
improving over the baselines by margins ranging from 17\% to 111\%.

Moreover, \textit{white-box approaches, such as \tool and \openia, demonstrate stability across languages}, maintaining their performance within a relatively narrow F1-score range. In contrast, black-box models, like CodeBERT and CodeT5+, show greater variability, with performance dropping significantly for certain languages. 
For example, while CodeT5+ performs competitively on TypeScript (F1-score of 0.82), its performance declines sharply to just 0.50 on Python. 
The reason is that black-box approaches operate on final code outputs; thus, they could be sensitive to language-specific characteristics such as keywords, syntax, and conventions. Meanwhile, white-box approaches leverage LLMs' internal representations, allowing them to capture deeper semantic understanding rather than relying on shallow surface-level features such as token sequences.
As a result, the white-box approaches, like \tool or \openia, can exhibit greater robustness across languages and programming paradigms.
\subsubsection{Impact of Training Data Size}

\begin{figure}
    \centering
    \includegraphics[width=\columnwidth]{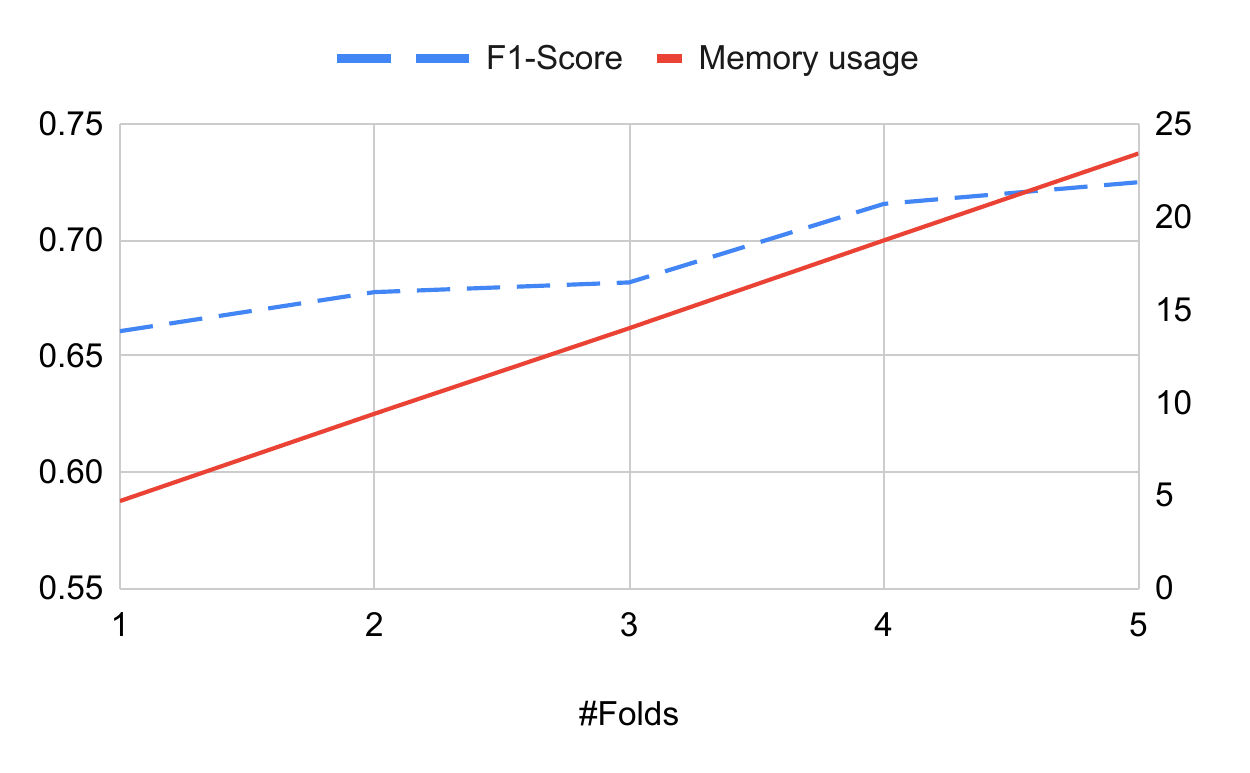}
    \caption{Impact of training size on \tool's performance and memory usage, left axis: \textit{F1-Score}; right axis: \textit{Memory Usage (GB)}}
    \label{fig:training_size}
\end{figure}

Figure~\ref{fig:training_size} shows that \textit{increasing the training data size enhances classification accuracy, but also leads to higher memory consumption.} Specifically, as the training size increases from 1 fold to 5 folds, the F1-score steadily improves from 0.66 to 0.72. This suggests that a larger training set enables the probing classifier and attention mechanism to better learn and generalize correctness-related patterns. 
Moreover, memory consumption grows almost linearly with the training size, ranging from around 5 GB at 1 fold to about 23 GB at 5 folds of data.
This is expected, as more training examples lead to larger sets of internal representations being processed and stored during training.
The results emphasize the importance of selecting a training size that balances performance needs with computational resources.
\subsubsection{Model Generalization Across Code LLMs}




\begin{figure}
    \centering
    \includegraphics[width=\columnwidth]{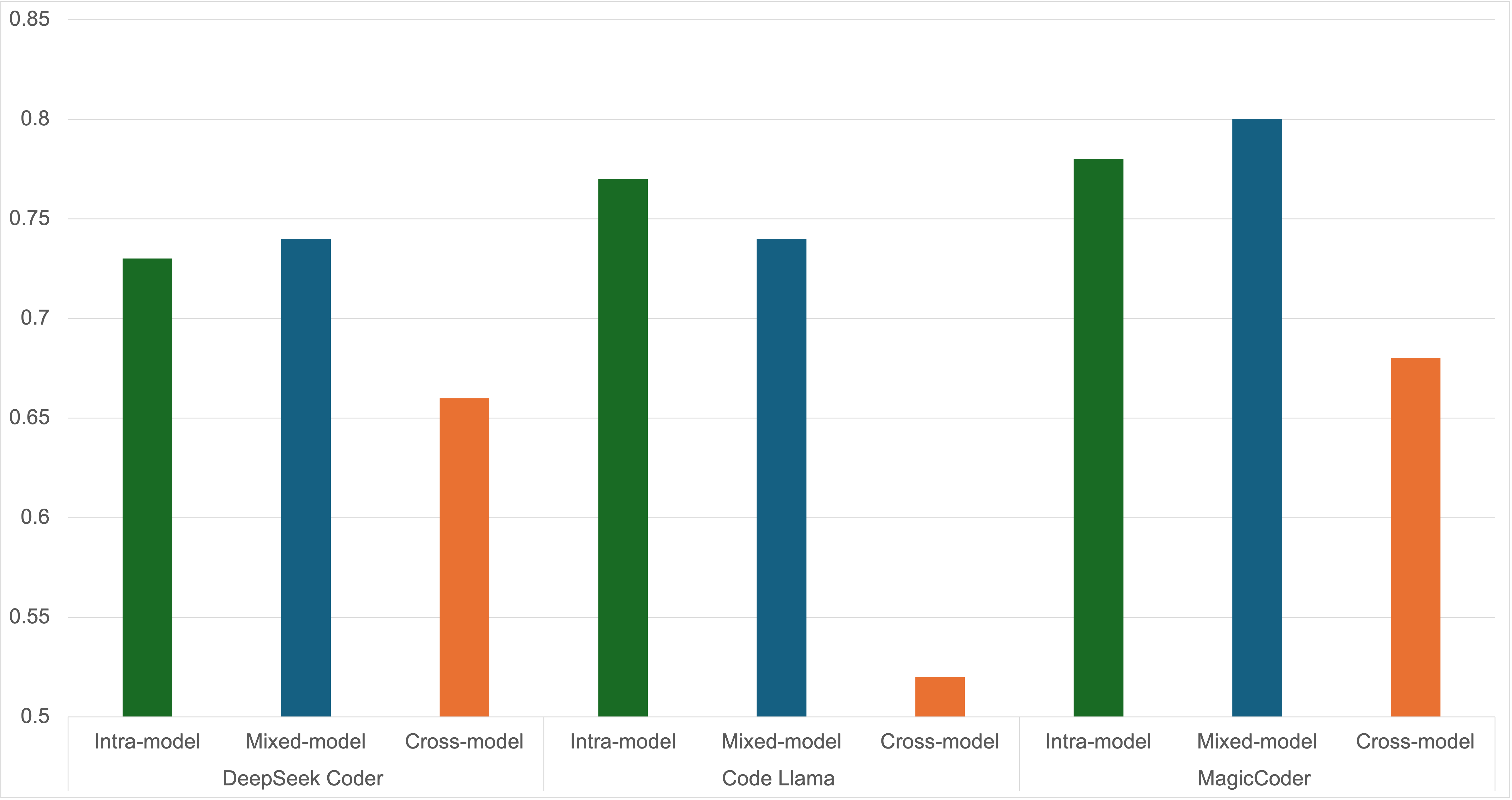}
    \caption{Impact of LLM generalization between training and inference on \tool's performance on F1-Score}
    \label{fig:sensivity_model}
\end{figure}

Figure~\ref{fig:sensivity_model} shows \tool's performance when trained on internal states from either the same or different models as the test-time model. This experiment employs three code LLMs, \deepseek-6.7B, \codellama-7B, and \magiccoder-7B, which share a common architecture of 32 layers and 4,096-dimensional hidden states. We evaluates three settings: \textit{intra-model}, \textit{cross-model}, and \textit{mixed-model}. In the \textit{intra-model} setting, the classifier is both trained and tested on the internal states of the same model. In the \textit{cross-model} setting, training is conducted on the internal states of two LLMs, while testing is performed on the remaining one. In the \textit{mixed-model} setting, the classifier is trained on the combined internal states of all three models and evaluated individually on each model.

Across all three models, \tool performs consistently better on \textit{\textit{intra-model} setting compared to on \textit{cross-model} setting}. For example, \tool on \textit{intra-model} for \codellama achieves an F1-Score of 0.77, whereas its performance on \textit{cross-model} results in a drop to 0.52. For \magiccoder and \deepseek, \tool's performance on \textit{cross-model} is also worse, achieving only 0.68 and 0.66, compared to its performance on \textit{intra-model}, over 0.78 and 0.73, respectively. 

The \textit{mixed-model} setting is designed to leverage both the specific characteristics of the test-time model and the generalizable patterns learned from the other models. However, as shown in Figure~\ref{fig:sensivity_model}, its effectiveness is not consistent across different target LLMs. For example, in the case of \magiccoder, \tool's performance on \textit{mixed-model} achieves a slightly higher F1-score than that on \textit{intra-model}. In contrast, for \codellama on \textit{mixed-model}, \tool performs worse than when it performs on \textit{intra-model}, indicating that the inclusion of external representations may introduce noise. These results suggest that its effectiveness depends on the degree of representational similarity among the models and the classifier's ability to extract shared correctness features.

Although \tool is model-agnostic, designed to dynamically select informative internal signals, the results suggest that correctness-related representations remain model-specific to some extent. Due to differences in architectural design, training dynamics, and pretraining data, the patterns of correctness-related signals might not be directly transferable across models. These findings highlight the importance of aligning the training and inference models to achieve optimal performance in white-box correctness assessment.

\begin{gtheorem} 
\textbf{Answer to RQ3}: \tool generalizes well across programming languages and data scales, with optimal performance achieved when training and inference use internal representations from the same code LLM.
\end{gtheorem}
\subsection{Efficiency Analysis}

The time complexity of \tool is primarily influenced by three components: (1) extracting internal representations, (2) training the attention scores and probing classifier, and (3) performing inference. 
Internal representations were extracted from the studied code LLMs on an NVIDIA A6000 GPU with 48GB of VRAM. This process required approximately 125 hours for the entire dataset. Importantly, the extraction was performed concurrently during code generation and did not introduce additional latency into the generation process.

All subsequent experiments, including training and inference of the probing classifier, were conducted on an Ubuntu Server 22.04  equipped with a single NVIDIA A6000 GPU (48GB VRAM) and 512GB of system RAM.
Since \tool bypasses costly embedding computations, its classifier training and inference are extremely fast. Specifically, the training phase requires just about 1 minute, and the inference phase takes only 0.6 milliseconds per generated code unit. Meanwhile, the black-box baselines such as CodeBERT and CodeT5+ are significantly more time-consuming, taking around 83 minutes to train, with over 98\% of the time spent on embedding generation.

Regarding memory usage, the standard variant of \tool consumes about 16.5 GB of GPU and CPU memory. However, 
When token position sampling is disabled, i.e., using the representations of all generated tokens, \tool's memory consumption increases substantially, reaching up to 300 GB.

\begin{gtheorem} 
\textbf{Answer to RQ4}: \tool achieves high efficiency in both training and inference. By employing token and layer sampling strategies it significantly reduces memory consumption without sacrificing performance.  This makes \tool a practical and scalable solution for real-world deployment.
\end{gtheorem}
\subsection{Threats to Validity}

The main threats to the validity of our work consist of internal, construct, and external threats.

\textbf{Internal Validity}. One threat to internal validity lies in the selection of hyperparameters. To mitigate this threat, we systematically explored various configurations of \tool and reused the official settings from the original papers for baselines~\cite{openia, codebert, codet5}, ensuring fair comparisons. Another potential threat involves the procedure for extracting internal representations. To address this threat, we relied on widely adopted PyTorch tools to extract hidden states during code generation. Additionally, to further reduce the risk of implementation errors, we carefully reviewed our code and made it public~\cite{website} so that other researchers can double-check and reproduce our experiments. 

\textbf{Construct Validity}. A threat to construct validity arises from the selection of evaluation metrics and experimental procedures. To mitigate this threat, we adopt standard  metrics, including Accuracy, Precision, Recall, and F1-Score for evaluation. To ensure a comprehensive assessment, we systematically evaluated \tool across multiple code LLMs, benchmarks, and experimental settings. Another potential threat involves the procedure for labeling code correctness, which relies on test outcomes or static analysis results. To mitigate this threat, we utilized high-quality, widely adopted benchmarks~\cite{HumanEval, MBPP, li-etal-2024-deveval, SecurityEval, Cweval, CODEGUARD+, Sallm} that include rigorous test suites. As future work, we plan to incorporate human evaluation to complement automated assessments.

\textbf{External Validity}.
External threat relates to the generalizability of our results. We mitigated this threat by evaluating \tool across diverse model families. However, due to hardware constraints, our evaluation focused on models with up to 13 billion parameters. We intend to explore the scalability of \tool to larger models in future work. 
Additionally, the prompt variability also poses a potential threat. To address this, we conducted evaluations on diverse benchmarks with varying prompt styles, offering partial mitigation. We also plan to systematically study prompt sensitivity in future experiments.
\section{Related Work}

\textbf{LLM-based code generation.}
LLMs, known for their powerful capabilities in contextual understanding and response generation, have been widely applied across various domains, including natural language processing~\cite{zhang2025systematic, li2024flexkbqa, wei2023empirical, zhang2025teleclass}, vision-language integration~\cite{zhang2024vision, chen2024vitamin}, and software engineering (SE) tasks~\cite{zheng2025towards, wang2025can, chen2024chatunitest, zhang2023repocoder, rambo}. 
In the context of SE, code LLMs, such as  \deepseek~\cite{deepseek-coder}, \codellama~\cite{codellama}, and \magiccoder~\cite{magicoder}, which have been pre-trained or fine-tuned on large-scale code corpora, have demonstrated remarkable success in a wide range of tasks, e.g. code generation~\cite{rambo, zhang2023repocoder, mu2024clarifygpt}, test generation~\cite{chen2024chatunitest, wang2024hits}, code summarization~\cite{sun2024source}, and program repair~\cite{jin2023inferfix}, etc.
These models are increasingly integrated into modern development environments via tools like GitHub Copilot~\cite{github_copilot} and CodeGeeX2~\cite{codegeex2}, where they assist developers in generating boilerplate code, completing functions, or writing test cases. 
Despite their impressive performance, a key challenge remains, the hallucination problem~\cite{zhang2025llm, liu2024exploring, liu2023your, wang2025towards}, where LLMs generate code that is syntactically valid but semantically incorrect, insecure, or unexecutable. Such hallucinations compromise the reliability of LLM-generated code and raise concerns about correctness, security, and trustworthiness. 
Mitigating such issues remains an open problem and is essential for ensuring the reliable and effective adoption of code LLMs in real-world software projects.

\textbf{LLM's hallucination detection.} Various approaches~\cite{ llmcheck, inside, inner-working,  manakul2023selfcheckgpt, huang2025survey} have been proposed to detect hallucinations in LLMs. These approaches can be broadly categorized into black-box and white-box approaches, depending on whether they rely on the model's input-output behaviors or leverage its internal computations.

\textbf{\textit{Black-box}} approaches~\cite{manakul2023selfcheckgpt, farquhar2024detecting, zhang2024self, zhang2024knowhalu, zhou2021detecting} detect hallucinations by relying solely on the input-output behavior of the models. A common strategy is \textit{uncertainty estimation}, based on a hypothesis that responses with higher model confidence are more likely to be correct~\cite{manakul2023selfcheckgpt, farquhar2024detecting, huang2023look}.
For instance, SelfCheckGPT~\cite{manakul2023selfcheckgpt} estimates model confidence by sampling multiple responses to the same query and measuring their semantic consistency. If the responses are consistent across samples, the output is considered to be reliable; otherwise, it is likely hallucinated.
Another research direction focuses on \textit{fact-checking}, where the generated outputs are validated against factual sources. These sources can be drawn from the model's internal knowledge~\cite{zhang2024self} or external corpora~\cite{zhang2024knowhalu, augenstein2024factuality, hu2024knowledge}. Additionally, some other black-box methods formulate hallucination detection as a \textit{classification task}, training a downstream classifier on features extracted from the generated outputs to distinguish hallucinated from faithful outputs~\cite{zhou2021detecting}.

On the other hand, \textbf{\textit{white-box}} approaches leverage the model's internal computations, such as \textit{hidden states}, \textit{activations}, \textit{attention maps}, to determine unreliable or hallucinated responses~\cite{inner-working, probing}. 
For example, Azaria~\etal~\cite{internal-state} train a classifier on the hidden activations to estimate the truthfulness of the generated statements. Similarly, INSIDE~\cite{inside} and FactoScope~\cite{factoscope}
analyze internal states to assess semantic consistency and factual reliability.

Our work follows the while-box paradigm by leveraging internal representations for hallucination detection. However, unlike prior studies~\cite{internal-state-2, inside, llmcheck, internal-state} that focus on unstructured Natural Language Generation (NLG) tasks, where correctness is often subjective by relying on human judgment or based on external factual knowledge, \tool target \textit{code correctness}, where correctness is objectively defined by syntax, semantics, and functional behaviors. 
Due to the nature of NLG tasks, previous works~\cite{inside, internal-state} mainly evaluate hallucination through semantic consistency or factual alignment among model responses. 
However, these techniques are not directly applicable to code.
In the context of code generation, a task may have multiple correct solutions that differ in implementation details or algorithms. As a result, inconsistency across generated responses does not necessarily imply incorrectness. 
To address this gap, \tool is specifically designed for assessing code correctness using internal representations.

\textbf{Code correctness assessment.}
Ensuring source code quality is a fundamental objective in the SE process. Traditional approaches for detecting program issues often rely heavily on handcrafted features and static analysis techniques~\cite{croft2021empirical}.
With the emergence of deep learning, pre-trained models such as CodeBERT~\cite{codebert}, GraphBERT~\cite{zhang2020graph}, and CodeT5~\cite{codet5} have been widely adopted due to their ability to capture rich contextual and semantic information from source code. 
These models support automatic feature extraction from different code representations such as raw code, program slices, or code property graphs, enabling bug and vulnerability detection across different levels of granularity, from coarse-grained levels, i.e., files or methods, to fine-grained~\cite{ivdetect, COSTA, velvet, linevd, linevul}, i.e., slices and statements. 
For instance, IVDetect~\cite{ivdetect} employs a graph-based neural network to detect vulnerabilities at the function level and uses interpretable models to further localize vulnerable statements. LineVul~\cite{linevul} and LineVD~\cite{linevd} leverage CodeBERT to encode code representations and have demonstrated superior performance over IVDetect in fine-grained vulnerability detection.

Recently, the quality of code generated by LLMs has gained significant attention due to the increasing reliance on AI-generated code in real-world applications. Several empirical studies have revealed the potential risks of bugs and security issues associated with LLM-generated code~\cite{lost-at-c, empirical-study-2, llm-gen-code-emp-study, calibration, vulnerabilities-copilot}. 
Multiple studies~\cite{huang2023look, llm-security-guard, autosafecoder} have been proposed for guaranteeing the quality of code generated by LLMs.
Huang \etal~\cite{huang2023look} propose a technique of uncertainty analysis to measure LLMs' confidence in generated outputs, which could aid in identifying potentially unreliable code.  
Furthermore, 
multi-agent frameworks such as AutoSafeCoder~\cite{autosafecoder} enhance LLM-based code generation by integrating static analysis and fuzz testing. Similarly, LLMSecGuard~\cite{llm-security-guard} combines the capabilities of static analyzers and LLMs to improve code security, demonstrating the benefits of hybrid approaches in mitigating vulnerabilities and strengthening code robustness.

\openia~\cite{openia} demonstrates that the internal states of LLMs encode valuable signals related to code correctness and can be leveraged to detect incorrect code. This work is closely related to ours, as both adopt a white-box approach. However, the key difference between \tool and \openia lies in the mechanism for selecting informative internal representations. 
\openia assumes that the hidden states of the last generated tokens at the last layer encapsulate the most comprehensive information and therefore uses them directly for correctness assessment. 
Instead of selecting a fixed representation that could lead to suboptimal performance across models, \tool proposes a model-agnostic approach that dynamically selects the most informative representations across multiple layers and token positions. This flexibility enables \tool to better adapt to different model architectures.
\section{Conclusion}

In this paper, we introduced \tool, a model-agnostic framework for assessing the correctness of LLM-generated code by dynamically selecting and leveraging informative internal representations. Specifically, \tool uses an attention-based mechanism to learn which internal states are most predictive of code correctness, enabling better adaptation across diverse LLM architectures and tasks.
Our experimental results demonstrate that \tool consistently outperforms SOTA black-box and white-box baselines across different correctness criteria, including compilability, functionality, and security. 
These findings highlight that dynamically selecting important internal signals enables \tool to serve as a robust and generalizable solution for assessing the correctness of code generated by various LLMs.

\printcredits

\bibliographystyle{elsarticle-num}
\bibliography{references}

\end{document}